\documentclass[aps,prb,manuscript]{revtex4}
\usepackage{amsmath,amssymb,epsfig}

\begin{document}
  
%%%%%%%%%%%%%%%%%%%%%%%%%%%%%%%%%%%%%%%%%%%%%%%%%%%%%%%%%%%%%%%%%%%%%%%%%%
\title{Ferromagnetism without flat bands in thin armchair nanoribbons}

\author{R\'eka~Trencs\'enyi and Zsolt~Gul\'acsi}
%$^{a,b}$, Arno~Kampf$^{a}$ 
%and Dieter~Vollhardt$^{a}$}
\address{
Department of Theoretical Physics, University of Debrecen, 
H-4010 Debrecen, Hungary}
\date{May 12, 2010}

\begin{abstract}
Describing by a Hubbard type of model a thin armchair graphene ribbon in the 
armchair hexagon chain limit,
one shows in exact terms, that even if the system does not have flat bands at all,
at low concentration a mesoscopic sample can have ferromagnetic ground state, being
metallic in the same time. The mechanism is connected to a common effect of 
correlations and confinement.
\end{abstract}
 \maketitle

%%%%%%%%%%%%%%%%%%%%%%%%%%%%%%%%%%%%%%%%%%%%%%%%%%%%%%%%%%%%%%%%%%%%%%%%%%%%
\section{Introduction}

Carbon-based nanoscale structures holding hexagonal repeating units and different type of
boundaries attract considerable interest by unprecedented application possibilities
in the design of organic nanodevices \cite{Be4,Be6}.
From these, nanoribbons are of
particular interest \cite{Be8}, being studied especially because of the emergence
possibilities of itinerant ferromagnetism \cite{Be15,Be17,Be18}, 
this subject being also driven
by the aim to produce magnetic behavior in organic materials not containing magnetic
elements \cite{Be13}. 

In these systems the shape of the edges is either of zig-zag or armchair type  
\cite{Be9}. The former case is 
usually associated to localized edge states \cite{Be14} which, due to 
their high degeneracy, can lead to flat-band ferromagnetism \cite{Be15} 
(in most cases a spin-polarized many particle, but localized ground state). 
The localized edge states for zig-zag edges on thick samples have been 
observed experimentally by scanning tunneling microscopy \cite{Be16}, but it is 
also known 
that in the thin 
ribbon case (the chain limit) the localized ferromagnetic nature is no more present 
in exact terms, and the spin polarized state becomes conducting in the low 
concentration limit for mesoscopic samples \cite{TKG}. Nevertheless, the emergence 
possibility of the edge localized states for zig-zag boundaries directed the attention 
to the armchair edges in the search for itinerant ferromagnetism in these structures.  

For armchair carbon-based nanoribbons the bulk states and the localized end states are
considered entangled by the short-range Coulomb repulsion \cite{Be15}. 
In these conditions, at the level of theoretical predictions,
itinerant carriers in dispersive bands are supposed to mediate via exchange
coupling the interaction among the local magnetic moments present in flat bands 
\cite{Be17,Be18}. Different approximate techniques are used for the description as: 
effective field theories combined with variational wave function approach \cite{Be17},
or analytical weak-coupling analysis combined with numerical density matrix 
renormalization-group method and first-principles calculations \cite{Be18}.
As a general observation one notes that the deduced results 
attract the attention to the 
importance of the Coulomb interaction and correlation effects in providing 
the physical properties of the systems under study \cite{Be15,Be18,Be18a}. This 
information is
important also in a broader context, since for example, potential nanoribbon 
applications in building up organic transistors or spin qubits \cite{Be19,Be20} 
do not take into account correlation effects at all. 

On the theoretical side the
studies directed to the explanation of ferromagnetism are in a stage 
present at other organic systems as well, for example pentagon chains \cite{Be13}, 
where the initial attempts based on flat band ferromagnetism
\cite{MT} encounter the difficulties of this model. One of these is
the fact that a perfectly flat band is difficult to obtain. The second difficulty is
related to the observation that the connectivity condition needed for the flat 
band ferromagnetism, is often not satisfied in experimentally manageable
situations \cite{Be21}, and this case is encountered for armchair nanoribbons
as well \cite{Be18}. The corrected theory in the light of this new input is usually
a some kind of developed flat band theory which 
does not need the in-flat-band connectivity
condition. On this line, for example in the case of the pentagon chains, a model 
applied in
describing rare earth compounds has been suggested to work \cite{Be21} using strong
hybridization effects \cite{Be22}, while in the
case of the armchair nanoribbons constructed from hexagon cells,
electrons from dispersive bands are 
conjectured to interconnect the moments created inside the flat bands \cite{Be17,Be18}. 
In fact, this last possibility is also known in the literature. In $D=1$ one 
dimensional 
case it originates from Ref. \cite{Be23}, whose extended version to $D > 1$ can be find
in Ref. \cite{Be24}. 

All the previously mentioned results relating ferromagnetism in organic 
nanoribbons suggest that spin-polarized states in such systems are intimately connected
to bare flat bands. This suggestion focuses the research directed to the understanding 
of
ferromagnetism (and this is the case also for graphene structures) 
exclusively on flat band emergence possibilities, in
condition when it is known that systems built up from hexagonal cells containing
several sublattices, usually 
do not contain flat bands \cite{RMP,TKG},
and flat bands are possible to appear only in special conditions, for example 
as consequences of edge states. Given by the importance of the application 
possibilities of
such organic systems in various fields of nanotechnology, this state of facts naturally 
leads to a main question: are indeed flat bands necessary for the emergence of 
ferromagnetism in structures built up from repeating hexagonal units ?
Guided by this question, with 
the aim to provide
relevant information and advancement in this field, one shows below in exact terms, 
that the answer to this question is negative.

In order to do that, one chooses the simplest possible armchair nanoribbon, an armchair
hexagon chain (see Fig.1) which besides the fact that represents a real existing
structure (i.e. polyphenanthrene), represents as well an armchair nanoribbon in the 
extreme thin limit. One uses for it a realistic Hubbard type of model which takes into
account as well the non-zero next nearest neighbor hopping amplitudes \cite{RMP}, and 
different on-site potentials on different type of sites. Even if the system 
is not integrable, using a special technique one deduces exact ground states for it. 
These, in the small concentration limit turn out to be itinerant and ferromagnetic in 
conditions in which flat bands are completely missing from the system. One shows that 
the spin polarization emerging for mesoscopic samples is a consequence of the common 
effect of correlations and confinement. Since increasing the ribbon width there is not
present a physical reason to cancel this behavior, we expect similar effect to occur
for thicker ribbons as well.

The method one uses is related to positive semidefinite operator properties. 
Based on these,
one transforms the Hamiltonian of the system ($\hat H$) in positive semidefinite form, 
the ground state being provided by the most general wave vector which gives the minimum 
possible eigenvalue (i.e. zero) of the positive semidefinite operators building up 
$\hat H$. 
The technique itself works well even in unexpected circumstances from the traditional 
point
of view of exact solutions as: three dimensions \cite{GV}, disordered and interacting
systems in two dimensions \cite{GD}, stripes, checkerboards and droplets in two dimensions
\cite{GM}, delocalization effect of the Hubbard interaction in two dimensions \cite{GIM}, 
ferroelectric systems \cite{fe1},
superconductors \cite{fe2,fe3}, or non-integrable
chain structures even under the action of external magnetic fields \cite{GKV1,GKV2}.
Different procedures of the method can be find in Refs.\cite{g1,g2,g3}, and extreme 
details
regarding its application in the case of the chain structures have been presented in
Refs. \cite{GKV1,GKV2}. 

One notes that the first exact results for the armchair hexagon chain in the interacting
case are contained in this paper. In order to deduce these, a special technique able to 
handle the
emerging extended operators in the presented case has been developed, used, 
and described. 
 
The remaining part of the paper is organized as follows. Section II. presents in 
details the 
studied system, Sect. III. describes the transformation of the Hamiltonian in positive
semidefinite form, Sect. IV. deduces the ground state wave function and presents the 
physical
properties of the obtained ferromagnetic phase, Sect. V. containing the summary and 
conclusions closes the presentation, and the Appendices A-C contain mathematical details.

\section{The system}

The studied system is an armchair hexagon chain whose cell constructed at the site
${\bf j}$ is presented in Fig.1. One has 8 in-cell positions at the sites
${\bf j}+{\bf r}_{\nu}$, $\nu=1,2,...,8$, ${\bf a}$ represents the Bravais 
vector of the chain, and for mathematical convenience one has ${\bf r}_1=0$.
One further underlines that the $\nu$ index
denotes as well eight different sublattices $S_{\nu}$ present in the system.

The Hamiltonian of the system can be written as
\begin{eqnarray}
\hat H=\hat T_0+\hat T_1+\hat T_2+\hat H_U,
\label{Keq1}
\end{eqnarray}
where the kinetic part $\hat H_0= \hat T_0+\hat T_1+\hat T_2$ contains the contributions
of the one-particle local potentials ($\hat T_0$), the nearest neighbor 
($\hat T_1$) and the next nearest neighbor ($\hat T_2$) hopping terms, while $\hat H_U$
represents the Hubbard interaction. One has 
%%%%%%%%%%%%%%%%%%%%%%%%%%%%%%%%%%%%%%%%%%%%%%%%%%%%%%%%%%%%%%%%%%%%%%%%%%
%% FIGURE 1
%%%%%%%%%%%%%%%%%%%%%%%%%%%%%%%%%%%%%%%%%%%%%%%%%%%%%%%%%%%%%%%%%%%%%%%%%%
\begin{figure}
\centerline{\includegraphics[width=12 cm,height=6 cm]{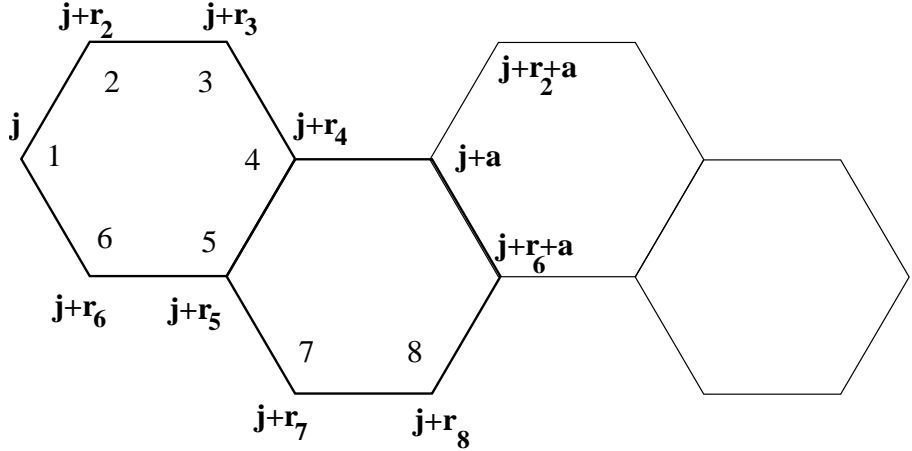}}
%csomopontok.eps}}
%\centerline{\epsfbox{Reka2.eps}}
\caption{The armchair hexagonal chain with cell constructed at the site ${\bf j}$ 
(thick line).
The in-cell positions are ${\bf j}+{\bf r}_{\nu}$, where $\nu=1,2,..8$ is 
denoting as well eight different sublattices. ${\bf a}$ is the Bravais vector, and one 
has ${\bf r}_1=0$. The numbers inside the cell represent the 
$\nu$ index of the in-cell sites.} 
\label{Kfig1}
\end{figure}
%%%%%%%%%%%%%%%%%%%%%%%%%%%%%%%%%%%%%%%%%%%%%%%%%%%%%%%%%%%%%%%%%%%%%%%%%%
\begin{eqnarray}
\hat T_0 &=& \sum_{\sigma}\sum_{n=0}^{N_c-1} \{\epsilon_0 \:[
\hat n_{{\bf j},\sigma} +
\hat n_{{\bf j}+{\bf r}_4,\sigma} +
\hat n_{{\bf j}+{\bf r}_5,\sigma} +
\hat n_{{\bf j}+{\bf r}_6,\sigma}]
\nonumber\\
&+& \epsilon_1 \:[
\hat n_{{\bf j}+{\bf r}_2,\sigma} +
\hat n_{{\bf j}+{\bf r}_3,\sigma} +
\hat n_{{\bf j}+{\bf r}_7,\sigma} +
\hat n_{{\bf j}+{\bf r}_8,\sigma}]\},
\nonumber\\
\hat T_1 &=& \sum_{\sigma}\sum_{n=0}^{N_c-1}\{t_1 \:[
\hat c_{{\bf j},\sigma}^{\dagger}\hat c_{{\bf j}+{\bf r}_6,\sigma} +
\hat c_{{\bf j}+{\bf r}_5,\sigma}^{\dagger} \hat c_{{\bf j}+{\bf r}_4,\sigma} + H.c.]
\nonumber\\
&+& t_2 \:[
\hat c_{{\bf j}+{\bf r}_6,\sigma}^{\dagger} \hat c_{{\bf j}+{\bf r}_5,\sigma} +
\hat c_{{\bf j}+{\bf a},\sigma}^{\dagger} \hat c_{{\bf j}+{\bf r}_4,\sigma} + H.c.]
\nonumber\\
&+& t_3 \:[
\hat c_{{\bf j}+{\bf r}_2,\sigma}^{\dagger} \hat c_{{\bf j},\sigma} +
\hat c_{{\bf j}+{\bf r}_4,\sigma}^{\dagger} \hat c_{{\bf j}+{\bf r}_3,\sigma} +
\hat c_{{\bf j}+{\bf r}_7,\sigma}^{\dagger} \hat c_{{\bf j}+{\bf r}_5,\sigma} +
\hat c_{{\bf j}+{\bf r }_6 +{\bf a},\sigma}^{\dagger}\hat c_{{\bf j}+{\bf r}_8,\sigma} 
+ H.c.]
\nonumber\\
&+& t_4 \:[
\hat c_{{\bf j}+{\bf r}_3,\sigma}^{\dagger} \hat c_{{\bf j}+{\bf r}_2,\sigma} + 
\hat c_{{\bf j}+{\bf r}_8,\sigma}^{\dagger} \hat c_{{\bf j}+{\bf r}_7,\sigma} + H.c.]\},
\nonumber\\
\hat T_2 &=& \sum_{\sigma}\sum_{n=0}^{N_c-1}\{t'_1 \:[
\hat c_{{\bf j}+{\bf r}_3,\sigma}^{\dagger} \hat c_{{\bf j},\sigma} +
\hat c_{{\bf j}+{\bf r}_4,\sigma}^{\dagger} \hat c_{{\bf j}+{\bf r}_2,\sigma} +
\hat c_{{\bf j}+{\bf r}_5,\sigma}^{\dagger} \hat c_{{\bf j}+{\bf r}_3,\sigma} +
\hat c_{{\bf j}+{\bf r}_2,\sigma}^{\dagger} \hat c_{{\bf j}+{\bf r}_6,\sigma}
\nonumber\\
&+& \hat c_{{\bf j}+{\bf r}_4,\sigma}^{\dagger} \hat c_{{\bf j}+{\bf r}_7,\sigma} 
 +  \hat c_{{\bf j}+{\bf r}_5,\sigma}^{\dagger} \hat c_{{\bf j}+{\bf r}_8,\sigma}
 +  \hat c_{{\bf j}+{\bf r}_7,\sigma}^{\dagger} \hat c_{{\bf j}+{\bf r}_6+{\bf a},
\sigma}
\nonumber\\
&+& \hat c_{{\bf j}+{\bf r}_8,\sigma}^{\dagger} \hat c_{{\bf j}+{\bf a},\sigma}
 +  \hat c_{{\bf j}+{\bf a},\sigma}^{\dagger} \hat c_{{\bf j}+{\bf r}_3,\sigma}
 +  \hat c_{{\bf j}+{\bf r}_2+{\bf a},\sigma}^{\dagger} \hat c_{{\bf j}+{\bf r}_4,
\sigma}
\nonumber\\
&+& \hat c_{{\bf j}+{\bf r}_7,\sigma}^{\dagger} \hat c_{{\bf j}+{\bf r}_6,\sigma}
+  \hat c_{{\bf j}+{\bf r}_5+{\bf a},\sigma}^{\dagger} \hat c_{{\bf j}+{\bf r}_8,
\sigma} + H.c.]
\nonumber\\
&+& t'_2 \:[
\hat c_{{\bf j},\sigma}^{\dagger} \hat c_{{\bf j}+{\bf r}_5,\sigma} +
\hat c_{{\bf j}+{\bf r}_6,\sigma}^{\dagger} \hat c_{{\bf j}+{\bf r}_4,\sigma} +
\hat c_{{\bf j}+{\bf a},\sigma}^{\dagger} \hat c_{{\bf j}+{\bf r}_5,\sigma} +
\hat c_{{\bf j}+{\bf r}_6+{\bf a},\sigma}^{\dagger} \hat c_{{\bf j}+{\bf r}_4,
\sigma} + H.c.]\},
\nonumber\\
\hat H_U &=& \sum_{n=0}^{N_c-1}\{U_0 \:[
\hat n_{{\bf j},\sigma}\hat n_{{\bf j},-\sigma} + 
\hat n_{{\bf j}+{\bf r}_4,\sigma} \hat n_{{\bf j}+{\bf r}_4,-\sigma} + 
\hat n_{{\bf j} +{\bf r}_5,\sigma} \hat n_{{\bf j}+{\bf r}_5,-\sigma} + 
\hat n_{{\bf j}+{\bf r}_6,\sigma} \hat n_{{\bf j}+{\bf r}_6,-\sigma}]
\nonumber\\
&+& U_1 \:[
\hat n_{{\bf j}+{\bf r}_2,\sigma} \hat n_{{\bf j}+{\bf r}_2,-\sigma} +
\hat n_{{\bf j}+{\bf r}_3,\sigma} \hat n_{{\bf j}+{\bf r}_3,-\sigma} +
\hat n_{{\bf j}+{\bf r}_7,\sigma} \hat n_{{\bf j}+{\bf r}_7,-\sigma} +
\hat n_{{\bf j}+{\bf r}_8,\sigma} \hat n_{{\bf j}+{\bf r}_8,-\sigma}]\}.
\label{Keq2}
\end{eqnarray}
In (\ref{Keq2}) $\hat c^{\dagger}_{{\bf i},\sigma}$ creates an electron with spin
$\sigma$ at the site ${\bf i}$, $\hat n_{{\bf i},\sigma}$ is the particle
number operator, $N_c$ represents the number of cells, and during the summation over $n$,
${\bf j}={\bf i}+n {\bf a}$ is considered, ${\bf i}$ being an arbitrary site of the
sublattice $S_{\nu=1}$. The parameters of $\hat H$ describing a realistic system 
are presented in Fig.2. The nearest neighbor hoppings are $t_1$ (touching bonds between 
hexagons), $t_2$ (horizontal hopping on the armchair), $t_3$ (the oblique 
hopping on the
armchair), and $t_4$ (the hopping on the external armchair bond). The next nearest 
neighbor
hoppings are $t'_1$ (bonds with only one end on touching points between hexagons), 
and $t'_2$ (bonds with both ends on touching points between hexagons). 
Furthermore, one
notes that the one-particle on-site potentials ($\epsilon_{\alpha}$) and the on site 
Coulomb repulsions $U_{\alpha}$ are denoted by the index $\alpha=0$ on the touching 
points between hexagons, and by the index $\alpha=1$ on other 
(external) sites. For mathematical 
simplicity one considers $U_0=U_1=U > 0$, and periodic boundary conditions are taken 
into account.

%%%%%%%%%%%%%%%%%%%%%%%%%%%%%%%%%%%%%%%%%%%%%%%%%%%%%%%%%%%%%%%%%%%%%%%%%%
%% FIGURE 2
%%%%%%%%%%%%%%%%%%%%%%%%%%%%%%%%%%%%%%%%%%%%%%%%%%%%%%%%%%%%%%%%%%%%%%%%%%
\begin{figure}
\centerline{\includegraphics[width=12 cm,height=6 cm]{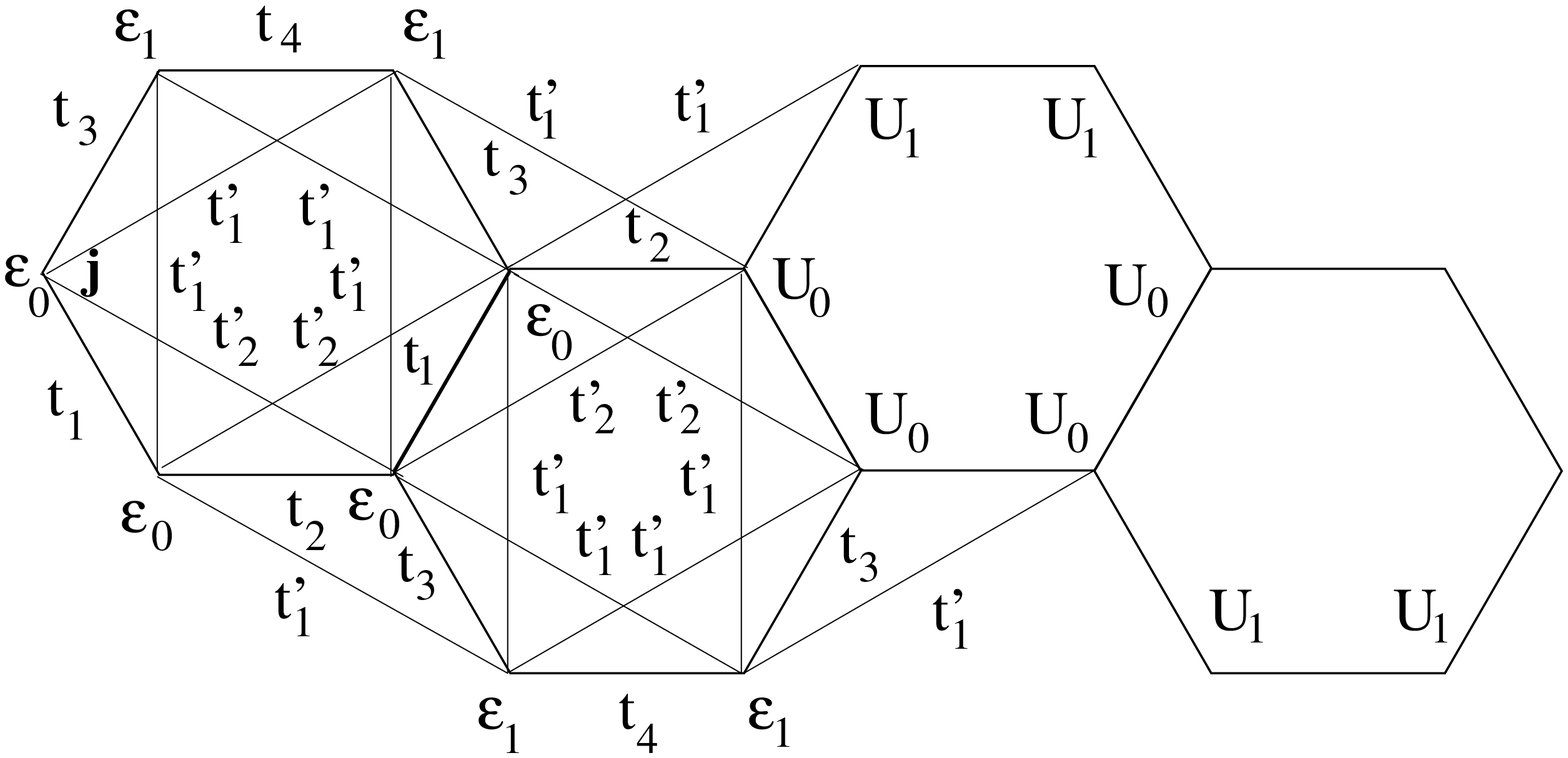}}
%\centerline{\epsfbox{Reka3.eps}}
\caption{The Hamiltonian parameters (see text following (\ref{Keq2})).} 
\label{Kfig2}
\end{figure}
%%%%%%%%%%%%%%%%%%%%%%%%%%%%%%%%%%%%%%%%%%%%%%%%%%%%%%%%%%%%%%%%%%%%%%%%%%

\section{The transformation of the Hamiltonian in a positive semidefinite form}

\subsection{The defined block operators}

In order to transform the Hamiltonian (\ref{Keq1},\ref{Keq2}) in a positive 
semidefinite form one defines eight block operators $\hat A_{p,{\bf j},\sigma}$,
$p=1,2,...,8$ for each site ${\bf j}$ of the sublattice $\nu=1$, as follows 

\begin{eqnarray}
\hat A_{1,{\bf j},\sigma} &=& 
a_1 \hat c_{{\bf j},\sigma} + 
a_2 \hat c_{{\bf j}+{\bf r}_2,\sigma} + 
a_3 \hat c_{{\bf j}+{\bf r}_3,\sigma} + 
a_4 \hat c_{{\bf j}+{\bf r}_4,\sigma}
\nonumber\\
\hat A_{2,{\bf j},\sigma} &=& 
b_1 \hat c_{{\bf j},\sigma} + 
b_4 \hat c_{{\bf j}+{\bf r}_4,\sigma} + 
b_5 \hat c_{{\bf j}+{\bf r}_5,\sigma} + 
b_6 \hat c_{{\bf j}+{\bf r}_6,\sigma}
\nonumber\\
\hat A_{3,{\bf j},\sigma} &=& 
d_0 \hat c_{{\bf j}+{\bf a},\sigma} + 
d_4 \hat c_{{\bf j}+{\bf r}_4,\sigma} + 
d_5 \hat c_{{\bf j}+{\bf r}_5,\sigma} + 
d_6 \hat c_{{\bf j}+{\bf r}_6+{\bf a},\sigma}
\nonumber\\
\hat A_{4,{\bf j},\sigma} &=& 
e_5 \hat c_{{\bf j}+{\bf r}_5,\sigma} + 
e_6 \hat c_{{\bf j}+{\bf r}_6+{\bf a},\sigma} + 
e_7 \hat c_{{\bf j}+{\bf r}_7,\sigma} + 
e_8 \hat c_{{\bf j}+{\bf r}_8,\sigma}
\nonumber\\
\hat A_{5,{\bf j},\sigma} &=& 
f_0 \hat c_{{\bf j}+{\bf a},\sigma} + 
f_3 \hat c_{{\bf j}+{\bf r}_3,\sigma} + 
f_4 \hat c_{{\bf j}+{\bf r}_4,\sigma} + 
f_5 \hat c_{{\bf j}+{\bf r}_5,\sigma}
\nonumber\\
\hat A_{6,{\bf j},\sigma} &=& 
g_0 \hat c_{{\bf j}+{\bf a},\sigma} + 
g_2 \hat c_{{\bf j}+{\bf r}_2+{\bf a},\sigma} + 
g_4 \hat c_{{\bf j}+{\bf r}_4,\sigma} + 
g_6 \hat c_{{\bf j}+{\bf r}_6+{\bf a},\sigma}
\nonumber\\
\hat A_{7,{\bf j},\sigma} &=& 
h_0 \hat c_{{\bf j}+{\bf a},\sigma} + 
h_5 \hat c_{{\bf j}+{\bf r}_5+{\bf a},\sigma} + 
h_6 \hat c_{{\bf j}+{\bf r}_6+{\bf a},\sigma} + 
h_8 \hat c_{{\bf j}+{\bf r}_8,\sigma}
\nonumber\\
\hat A_{8,{\bf j},\sigma} &=& 
k_4 \hat c_{{\bf j}+{\bf r}_4,\sigma} + 
k_5 \hat c_{{\bf j}+{\bf r}_5,\sigma} + 
k_6 \hat c_{{\bf j}+{\bf r}_6,\sigma} + 
k_7 \hat c_{{\bf j}+{\bf r}_7,\sigma} .
\label{Keq3}
\end{eqnarray}

\begin{figure}
\centerline{\includegraphics[width=7 cm,height=6 cm]{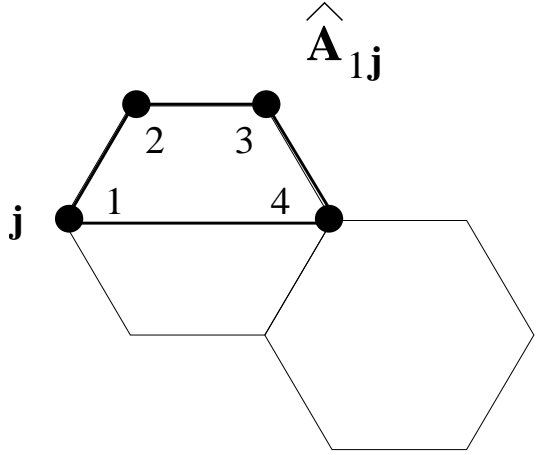}}
%\centerline{\epsfbox{Reka2.eps}}
\caption{The block $p=1$ depicted in the cell defined at the site ${\bf j}$ 
providing the block operator $\hat A_{1, {\bf j},\sigma}$. The thick line shows 
the block, while the numbers are indicating the $\nu$ index of the sites 
whose position is ${\bf j}+{\bf r}_{\nu}$.} 
\label{Kfig3}
\end{figure}

\begin{figure}
\centerline{\includegraphics[width=7 cm,height=6 cm]{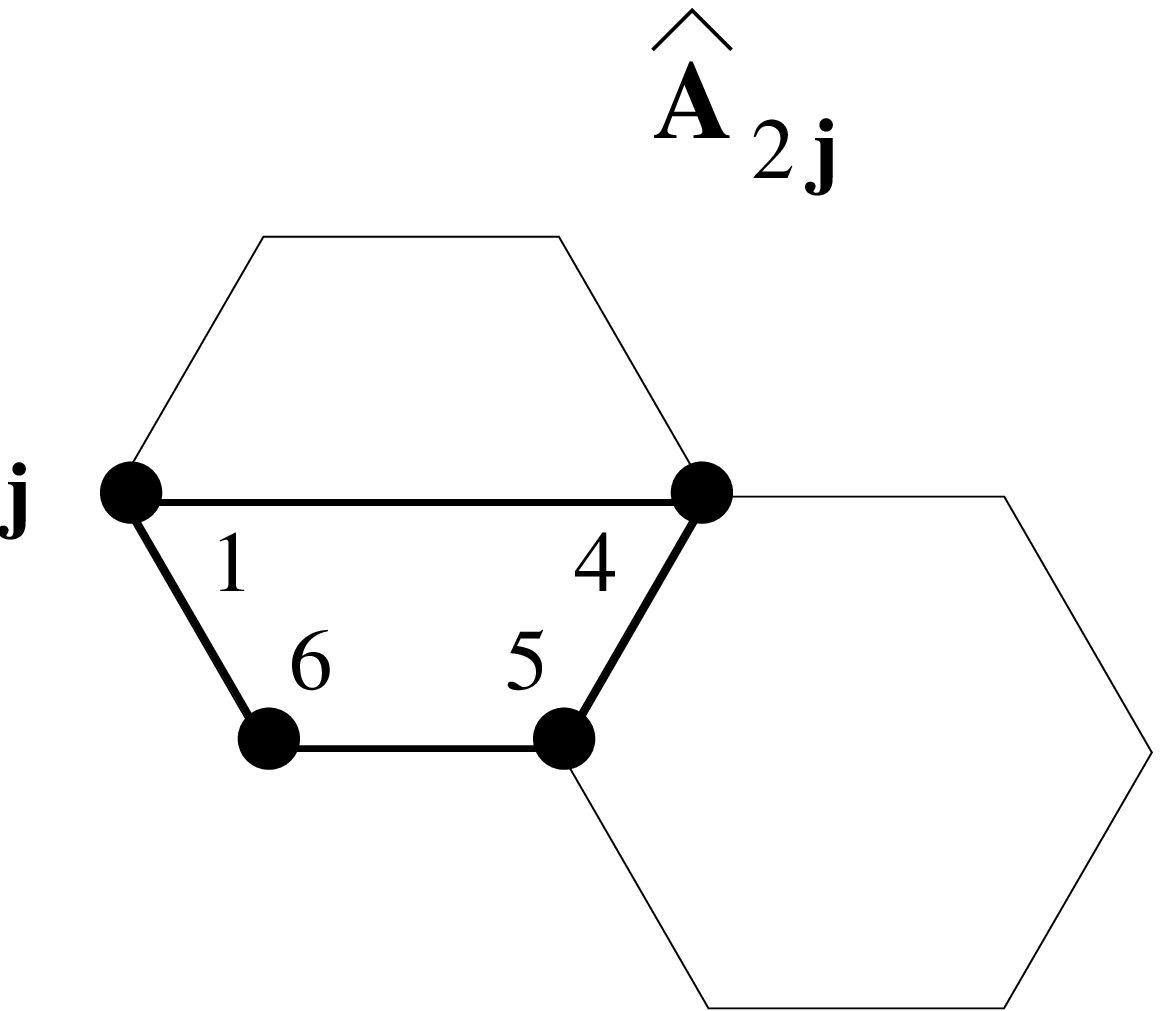}}
%\centerline{\epsfbox{Reka2.eps}}
\caption{The block $p=2$ depicted in the cell defined at the site ${\bf j}$ 
providing the block operator $\hat A_{2, {\bf j},\sigma}$. The thick line shows 
the block, while the numbers are indicating the $\nu$ index of the sites 
whose position is ${\bf j}+{\bf r}_{\nu}$.} 
\label{Kfig4}
\end{figure}

\begin{figure}
\centerline{\includegraphics[width=8 cm,height=6 cm]{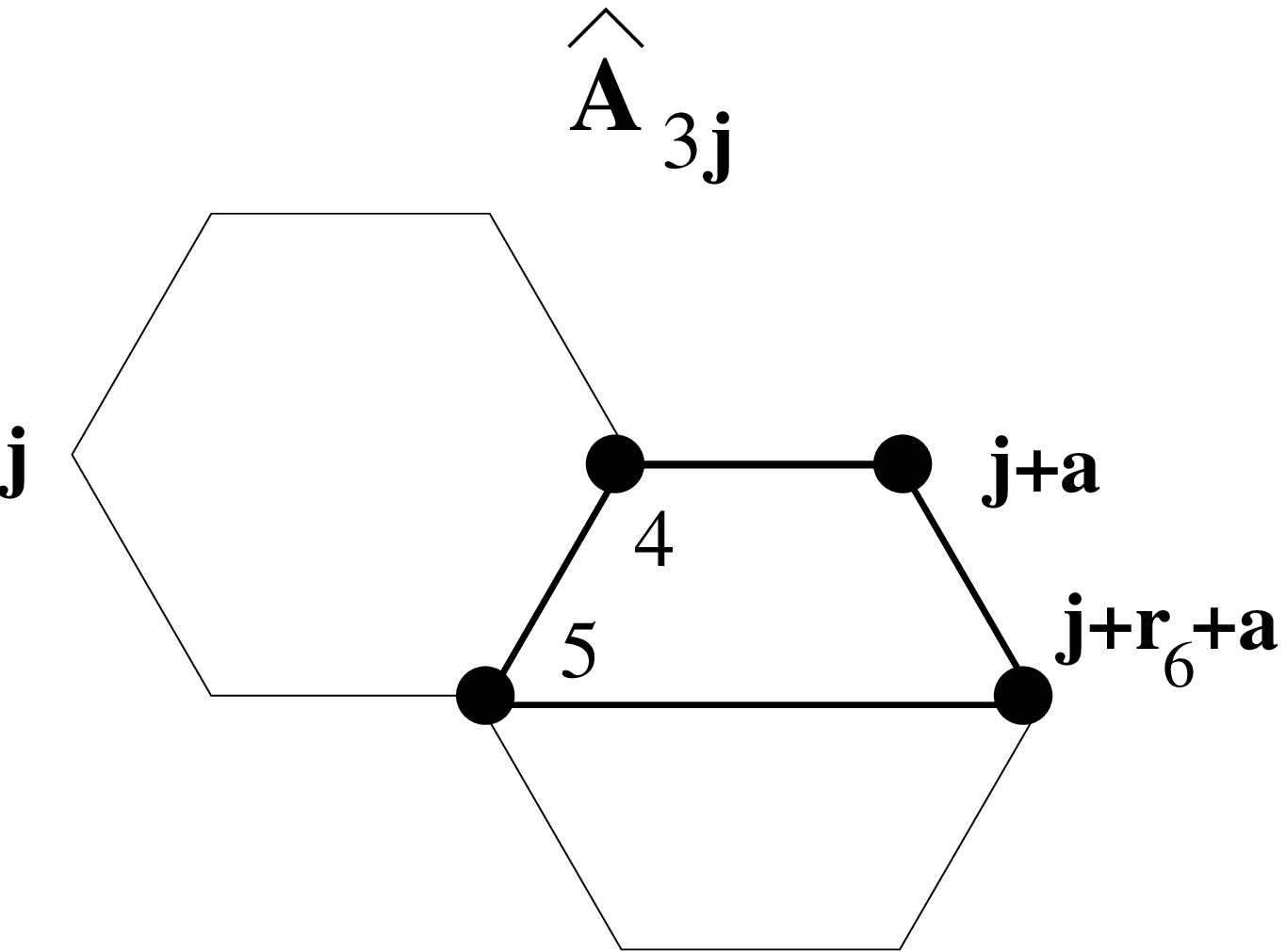}}
%\centerline{\epsfbox{Reka2.eps}}
\caption{The block $p=3$ depicted in the cell defined at the site ${\bf j}$ 
providing the block operator $\hat A_{3, {\bf j},\sigma}$. The thick line shows 
the block, while the numbers are indicating the $\nu$ index of the sites 
connected exclusively to a single cell.} 
\label{Kfig5}
\end{figure}

\begin{figure}
\centerline{\includegraphics[width=8 cm,height=6 cm]{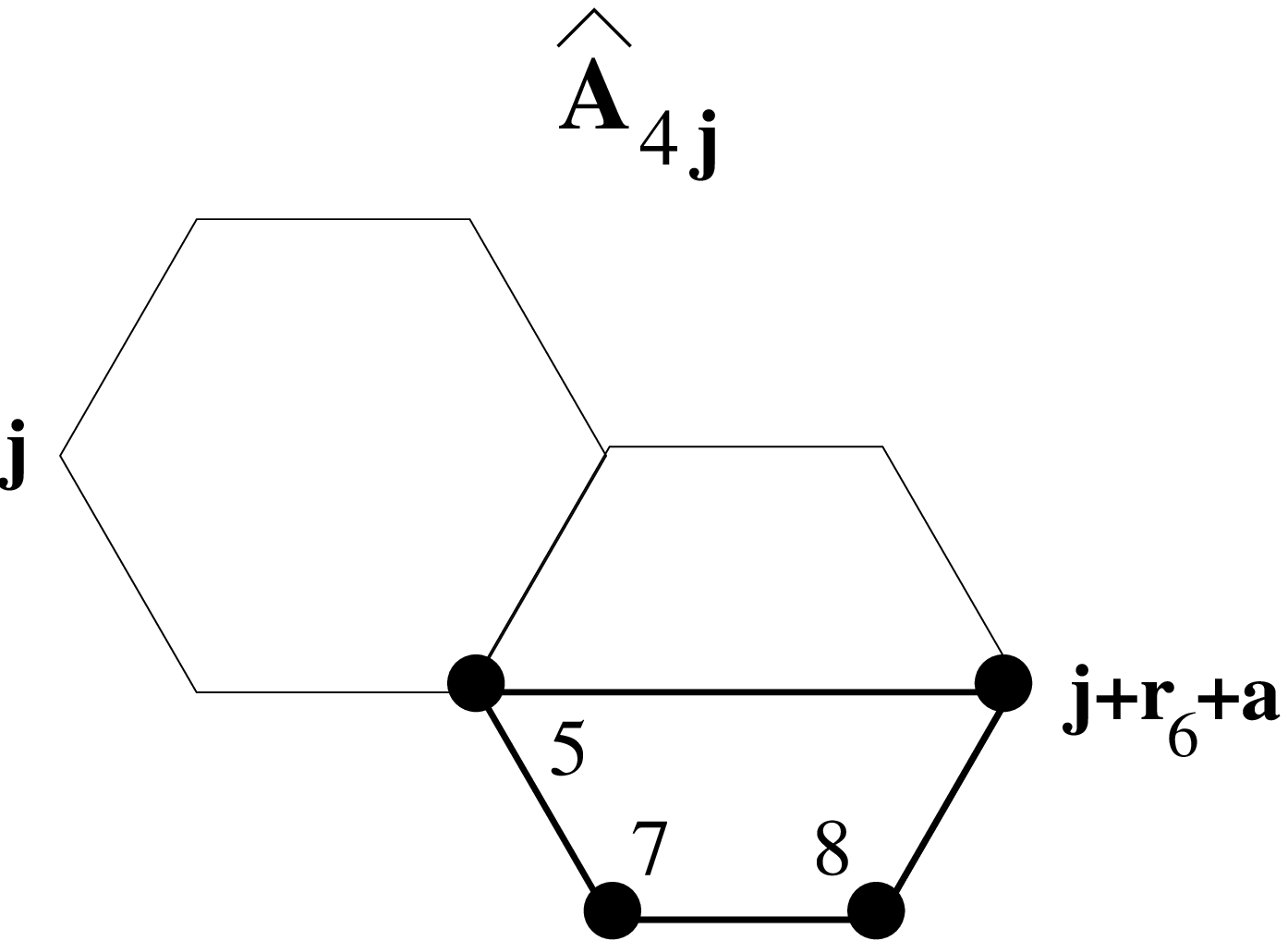}}
%\centerline{\epsfbox{Reka2.eps}}
\caption{The block $p=4$ depicted in the cell defined at the site ${\bf j}$ 
providing the block operator $\hat A_{4, {\bf j},\sigma}$. The thick line shows 
the block, while the numbers are indicating the $\nu$ index of the sites 
connected exclusively to a single cell.}
\label{Kfig6}
\end{figure}

\begin{figure}
\centerline{\includegraphics[width=8 cm,height=6 cm]{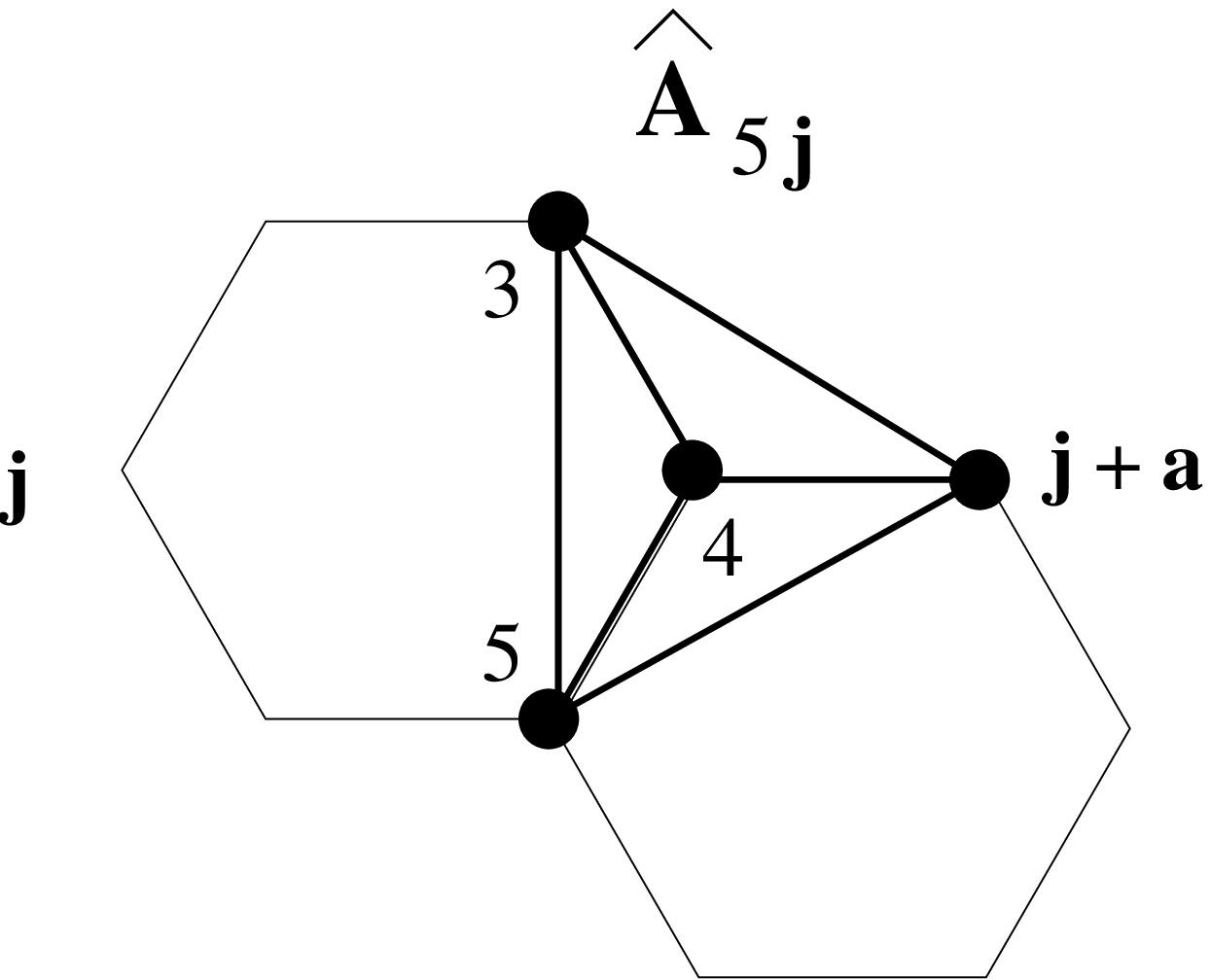}}
%\centerline{\epsfbox{Reka2.eps}}
\caption{The block $p=5$ depicted in the cell defined at the site ${\bf j}$ 
providing the block operator $\hat A_{5, {\bf j},\sigma}$. The thick line shows 
the block, while the numbers are indicating the $\nu$ index of the sites 
whose position is ${\bf j}+{\bf r}_{\nu}$.} 
\label{Kfig7}
\end{figure}

\begin{figure}
\centerline{\includegraphics[width=10 cm,height=6 cm]{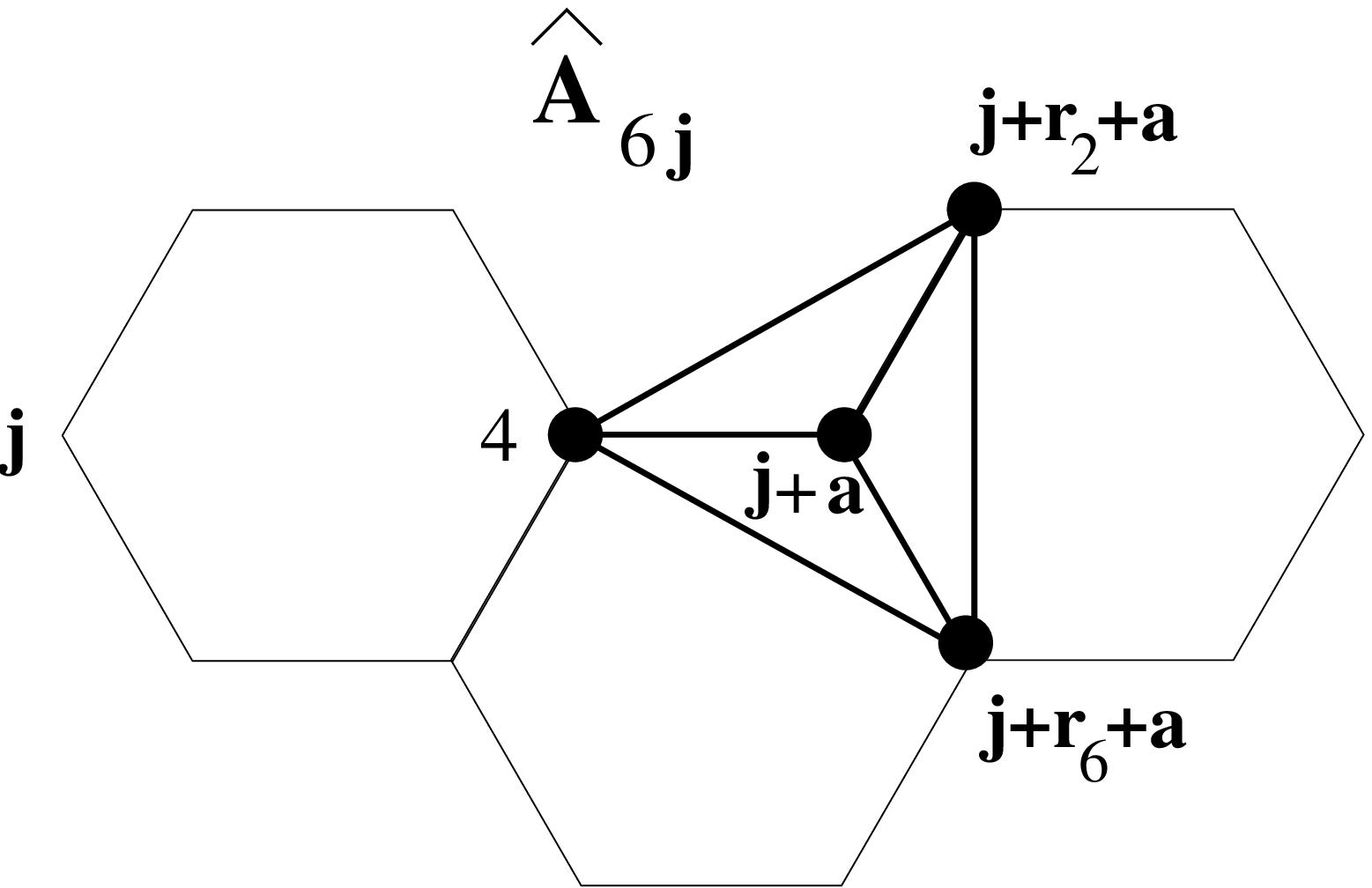}}
%\centerline{\epsfbox{Reka2.eps}}
\caption{The block $p=6$ depicted in the cell defined at the site ${\bf j}$ 
providing the block operator $\hat A_{6, {\bf j},\sigma}$. The thick line shows 
the block, the number four indicates the $\nu=4$ site. Other block sites (placed in the 
following cell) are indicated by their position
${\bf j}+{\bf r}_{\nu} +{\bf a}$.}  
\label{Kfig8}
\end{figure}

\begin{figure}
\centerline{\includegraphics[width=10 cm,height=6 cm]{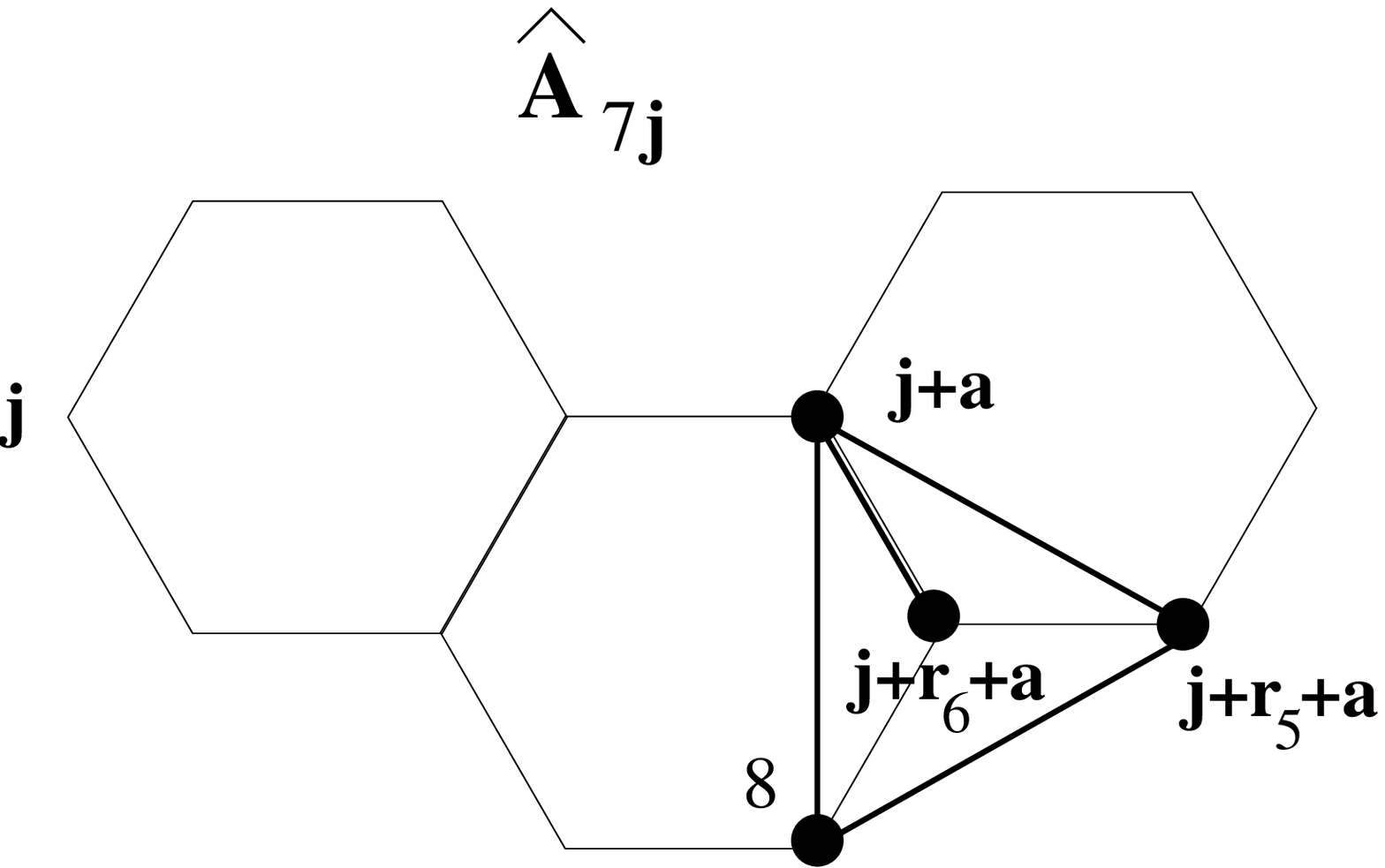}}
%\centerline{\epsfbox{Reka2.eps}}
\caption{The block $p=7$ depicted in the cell defined at the site ${\bf j}$ 
providing the block operator $\hat A_{7, {\bf j},\sigma}$. The thick line shows 
the block, the number eight indicates the $\nu=8$ site. Other block sites (placed in the 
following cell) are indicated by their position
${\bf j}+{\bf r}_{\nu} +{\bf a}$.}  
\label{Kfig9}
\end{figure}

\begin{figure}
\centerline{\includegraphics[width=7 cm,height=6 cm]{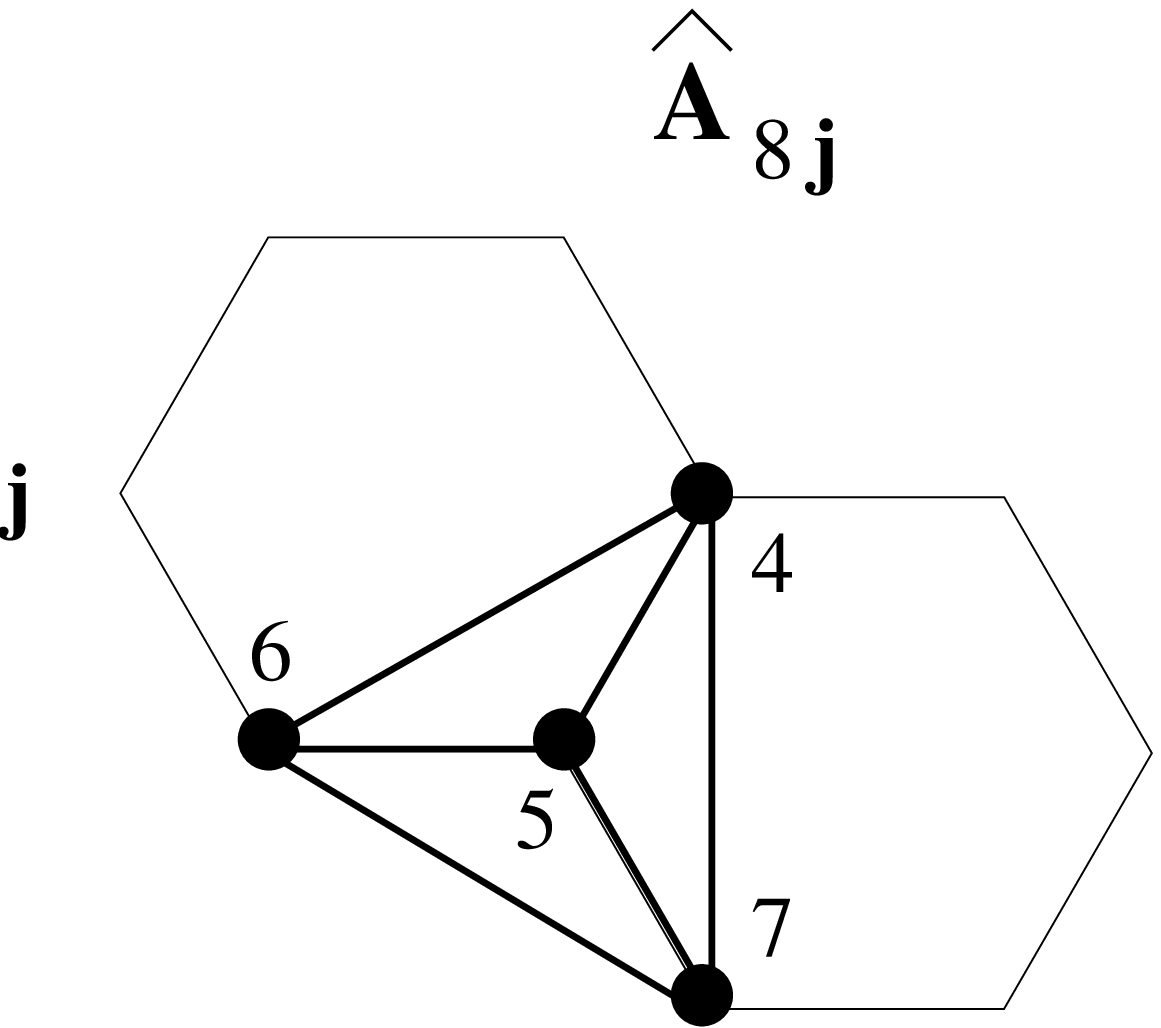}}
%\centerline{\epsfbox{Reka2.eps}}
\caption{The block $p=8$ depicted in the cell defined at the site ${\bf j}$ 
providing the block operator $\hat A_{8, {\bf j},\sigma}$. The thick line shows 
the block, while the numbers are indicating the $\nu$ index of the sites 
whose position is ${\bf j}+{\bf r}_{\nu}$.} 
\label{Kfig10}
\end{figure}

As  seen from (\ref{Keq3}) the used block operators $\hat A_{p,{\bf j},\sigma}$
are linear combinations of the starting fermionic operators $\hat c_{{\bf i},\sigma}$
acting on the sites of the block $p$ defined at the site ${\bf j}$ (see
Figs.\ref{Kfig3}-\ref{Kfig10}). The numerical prefactors 
$a_i, b_i,d_i,...,k_i$ are unknown at the moment, and will be deduced further on.

%%%%%%%%%%%%%%%%%%%%%%%%%%%%%%%%%%%%%%%%%%%%%%%%%%%%%%%%%%%%%%%%%%%%%%%%%%
%% FIGURE 3
%%%%%%%%%%%%%%%%%%%%%%%%%%%%%%%%%%%%%%%%%%%%%%%%%%%%%%%%%%%%%%%%%%%%%%%%%%
%%%%%%%%%%%%%%%%%%%%%%%%%%%%%%%%%%%%%%%%%%%%%%%%%%%%%%%%%%%%%%%%%%%%%%%%%%

%%%%%%%%%%%%%%%%%%%%%%%%%%%%%%%%%%%%%%%%%%%%%%%%%%%%%%%%%%%%%%%%%%%%%%%%%%
%% FIGURE 4
%%%%%%%%%%%%%%%%%%%%%%%%%%%%%%%%%%%%%%%%%%%%%%%%%%%%%%%%%%%%%%%%%%%%%%%%%%
%%%%%%%%%%%%%%%%%%%%%%%%%%%%%%%%%%%%%%%%%%%%%%%%%%%%%%%%%%%%%%%%%%%%%%%%%%

%%%%%%%%%%%%%%%%%%%%%%%%%%%%%%%%%%%%%%%%%%%%%%%%%%%%%%%%%%%%%%%%%%%%%%%%%%
%% FIGURE 5
%%%%%%%%%%%%%%%%%%%%%%%%%%%%%%%%%%%%%%%%%%%%%%%%%%%%%%%%%%%%%%%%%%%%%%%%%%
%%%%%%%%%%%%%%%%%%%%%%%%%%%%%%%%%%%%%%%%%%%%%%%%%%%%%%%%%%%%%%%%%%%%%%%%%%

%%%%%%%%%%%%%%%%%%%%%%%%%%%%%%%%%%%%%%%%%%%%%%%%%%%%%%%%%%%%%%%%%%%%%%%%%%
%% FIGURE 6
%%%%%%%%%%%%%%%%%%%%%%%%%%%%%%%%%%%%%%%%%%%%%%%%%%%%%%%%%%%%%%%%%%%%%%%%%%
%%%%%%%%%%%%%%%%%%%%%%%%%%%%%%%%%%%%%%%%%%%%%%%%%%%%%%%%%%%%%%%%%%%%%%%%%%

%%%%%%%%%%%%%%%%%%%%%%%%%%%%%%%%%%%%%%%%%%%%%%%%%%%%%%%%%%%%%%%%%%%%%%%%%%
%% FIGURE 7
%%%%%%%%%%%%%%%%%%%%%%%%%%%%%%%%%%%%%%%%%%%%%%%%%%%%%%%%%%%%%%%%%%%%%%%%%%
%%%%%%%%%%%%%%%%%%%%%%%%%%%%%%%%%%%%%%%%%%%%%%%%%%%%%%%%%%%%%%%%%%%%%%%%%%

%%%%%%%%%%%%%%%%%%%%%%%%%%%%%%%%%%%%%%%%%%%%%%%%%%%%%%%%%%%%%%%%%%%%%%%%%%
%% FIGURE 8
%%%%%%%%%%%%%%%%%%%%%%%%%%%%%%%%%%%%%%%%%%%%%%%%%%%%%%%%%%%%%%%%%%%%%%%%%%
%%%%%%%%%%%%%%%%%%%%%%%%%%%%%%%%%%%%%%%%%%%%%%%%%%%%%%%%%%%%%%%%%%%%%%%%%%

%%%%%%%%%%%%%%%%%%%%%%%%%%%%%%%%%%%%%%%%%%%%%%%%%%%%%%%%%%%%%%%%%%%%%%%%%%
%% FIGURE 9
%%%%%%%%%%%%%%%%%%%%%%%%%%%%%%%%%%%%%%%%%%%%%%%%%%%%%%%%%%%%%%%%%%%%%%%%%%
%%%%%%%%%%%%%%%%%%%%%%%%%%%%%%%%%%%%%%%%%%%%%%%%%%%%%%%%%%%%%%%%%%%%%%%%%%

%%%%%%%%%%%%%%%%%%%%%%%%%%%%%%%%%%%%%%%%%%%%%%%%%%%%%%%%%%%%%%%%%%%%%%%%%%
%% FIGURE 10
%%%%%%%%%%%%%%%%%%%%%%%%%%%%%%%%%%%%%%%%%%%%%%%%%%%%%%%%%%%%%%%%%%%%%%%%%%
%%%%%%%%%%%%%%%%%%%%%%%%%%%%%%%%%%%%%%%%%%%%%%%%%%%%%%%%%%%%%%%%%%%%%%%%%%

\subsection{The positive semidefinite form of the Hamiltonian}

Based on the block operators defined in (\ref{Keq3}), the starting Hamiltonian presented
in (\ref{Keq1},\ref{Keq2}) is transformed (i.e. is rewritten in exact terms) in the 
following positive 
semidefinite form
\begin{eqnarray}
\hat H = \sum_{\sigma} \sum_{p=1}^8 \sum_{{\bf j}=1}^{N_c}
\hat A_{p,{\bf j},\sigma}^{\dagger} \hat A_{p,{\bf j},\sigma} + \hat H_U + K\hat N,
\label{Keq4}
\end{eqnarray}
where $\hat N$ represents the operator of the total number of electrons and $K$ is a
constant. The numerical prefactors in (\ref{Keq3}) and the constant $K$ are expressed
by i) effectuating the products $\hat A_{p,{\bf j},\sigma}^{\dagger} 
\hat A_{p,{\bf j},\sigma}$ and the prescribed sums in the first term of (\ref{Keq4}),
and ii) equating the obtained result term by term to the expression of the starting
Hamiltonian (\ref{Keq1},\ref{Keq2}). The obtained equations connecting the parameters
of the starting Hamiltonian to the parameters of the block operators and the constant
$K$ in the transcribed Hamiltonian (\ref{Keq4}) are the following ones: The hopping 
terms become expressed as  
\begin{eqnarray}
t_{{\bf j}+{\bf r}_5,{\bf j}+{\bf r}_4} &=& t_1 
= b_5^*b_4 + d_5^*d_4 + f_5^*f_4 + k_5^*k_4, \quad
t_{{\bf j}+{\bf r}_1,{\bf j}+{\bf r}_6} = 
t_1 = b_1^*b_6 + g_0^*g_6 + h_0^*h_6 + d_0^*d_6,
\nonumber\\
t_{{\bf j}+{\bf r}_6,{\bf j}+{\bf r}_5} &=& t_2 = b_6^*b_5 + h_6^*h_5 + k_6^*k_5, \quad
t_{{\bf j}+{\bf r}_1+{\bf a},{\bf j}+{\bf r}_4} = t_2 = d_0^*d_4 + f_0^*f_4 + g_0^*g_4,
\nonumber\\
t_{{\bf j}+{\bf r}_2,{\bf j}+{\bf r}_1} &=& t_3 = a_2^*a_1 + g_2^*g_0, \quad
t_{{\bf j}+{\bf r}_4,{\bf j}+{\bf r}_3} = t_3 = a_4^*a_3 + f_4^*f_3, 
\nonumber\\
t_{{\bf j}+{\bf r}_6+{\bf a},{\bf j}+{\bf r}_8} &=& t_3 = e_6^*e_8 + h_6^*h_8, \quad
t_{{\bf j}+{\bf r}_7,{\bf j}+{\bf r}_5} = t_3 = e_7^*e_5 + k_7^*k_5, 
\nonumber\\
t_{{\bf j}+{\bf r}_3,{\bf j}+{\bf r}_2} &=& t_4 = a_3^*a_2, \quad 
t_{{\bf j}+{\bf r}_8,{\bf j}+{\bf r}_7} = t_4 = e_8^*e_7,
\quad
t_{{\bf j}+{\bf r}_3,{\bf j}+{\bf r}_1} = t'_1 = a_3^*a_1, 
\nonumber\\
%%%%%%%%%%%%%%%%%%%%%%%%%%%%%%%%%%%%%%%%%%%%%%%%%%%%%%%%%%%%%% 
t_{{\bf j}+{\bf r}_5,{\bf j}+{\bf r}_3} &=& t'_1 = f_5^*f_3, \quad
t_{{\bf j}+{\bf r}_4,{\bf j}+{\bf r}_2} = t'_1 = a_4^*a_2, \quad 
t_{{\bf j}+{\bf r}_2,{\bf j}+{\bf r}_6} = t'_1 = g_2^*g_6,
\nonumber\\
t_{{\bf j}+{\bf r}_8,{\bf j}+{\bf r}_1+{\bf a}} &=& t'_1 = h_8^*h_0, \quad 
t_{{\bf j}+{\bf r}_5,{\bf j}+{\bf r}_8} = t'_1 = e_5^*e_8, \quad
t_{{\bf j}+{\bf r}_4,{\bf j}+{\bf r}_7} = t'_1 = k_4^*k_7, 
\nonumber\\
t_{{\bf j}+{\bf r}_7,{\bf j}+{\bf r}_6} &=& t'_1 = k_7^*k_6, \quad
t_{{\bf j}+{\bf r}_7,{\bf j}+{\bf r}_6+{\bf a}} = t'_1 = e_7^*e_6, \quad 
t_{{\bf j}+{\bf r}_1+{\bf a},{\bf j}+{\bf r}_3} = t'_1 = f_0^*f_3, 
\nonumber\\
t_{{\bf j}+{\bf r}_2+{\bf a},{\bf j}+{\bf r}_4} &=& t'_1 = g_2^*g_4, \quad
t_{{\bf j}+{\bf r}_5+{\bf a},{\bf j}+{\bf r}_8} = t'_1 = h_5^*h_8, \quad
t_{{\bf j}+{\bf r}_6+{\bf a},{\bf j}+{\bf r}_4} = t'_2 = d_6^*d_4 + g_6^*g_4, 
\nonumber\\
t_{{\bf j}+{\bf r}_1,{\bf j}+{\bf r}_5} &=& t'_2 = b_1^*b_5 + h_0^*h_5, \quad 
t_{{\bf j}+{\bf r}_6,{\bf j}+{\bf r}_4} = t'_2 = b_6^*b_4 + k_6^*k_4, 
\nonumber\\ 
t_{{\bf j}+{\bf r}_1+{\bf a},{\bf j}+{\bf r}_5} &=& t'_2 = d_0^*d_5 + f_0^*f_5,
\nonumber\\                                                    
t_{{\bf j}+{\bf r}_6+{\bf a},{\bf j}+{\bf r}_5} &=& 0 = d_6^*d_5 + e_6^*e_5, \quad
t_{{\bf j}+{\bf r}_4,{\bf j}+{\bf r}_1} = 0 = a_4^*a_1 + b_4^*b_1,
\label{Keq5}
\end{eqnarray}
while the on-site potentials are given by
\begin{eqnarray}
\epsilon_{{\bf j}+{\bf r}_1} &=& \epsilon_0 = |a_1|^2 + |b_1|^2 + |d_0|^2 + 
|f_0|^2 + |g_0|^2 + |h_0|^2 + K,
\nonumber\\
\epsilon_{{\bf j}+{\bf r}_2} &=& \epsilon_1 = |a_2|^2 + |g_2|^2 + K,
\nonumber\\
\epsilon_{{\bf j}+{\bf r}_3} &=& \epsilon_1 = |a_3|^2 + |f_3|^2 + K,
\nonumber\\
\epsilon_{{\bf j}+{\bf r}_4} &=& \epsilon_0 = |a_4|^2 + |b_4|^2 + |d_4|^2 + 
|f_4|^2 + |g_4|^2 + |k_4|^2 + K,
\nonumber\\
\epsilon_{{\bf j}+{\bf r}_5} &=& \epsilon_0 = |b_5|^2 + |d_5|^2 + |e_5|^2 + 
|f_5|^2 + |h_5|^2 + |k_5|^2 + K,
\nonumber\\
\epsilon_{{\bf j}+{\bf r}_6} &=& \epsilon_0 = |b_6|^2 + |d_6|^2 + |e_6|^2 + 
|g_6|^2 + |h_6|^2 + |k_6|^2 + K,
\nonumber\\
\epsilon_{{\bf j}+{\bf r}_7} &=& \epsilon_1 = |e_7|^2 + |k_7|^2 + K,
\nonumber\\
\epsilon_{{\bf j}+{\bf r}_8} &=& \epsilon_1 = |e_8|^2 + |h_8|^2 + K .
\label{Keq6}
\end{eqnarray}     
For clarity, in (\ref{Keq5}) one specifies also the hopping matrix elements 
$t_{{\bf i}_1,{\bf i}_2}$, as the component of $t_{{\bf i}_1,{\bf i}_2}
\hat c^{\dagger}_{{\bf i}_1,\sigma} \hat c_{{\bf i}_2,\sigma}$ kinetic energy term
which leads to the respective equation. Again for clarity, at the level of the  
on-site potentials in (\ref{Keq6}), in the index of $\epsilon_{\bf i}$, the site ${\bf i}$
where the analyzed potential is considered is precisely indicated.

One notes that
Eqs.(\ref{Keq5},\ref{Keq6}) are the matching conditions of the solution since
the transformation from (\ref{Keq1},\ref{Keq2}) to (\ref{Keq4}) is valid only if 
Eqs.(\ref{Keq5},\ref{Keq6}) are satisfied.

\subsection{The explicit expression of the operators $\hat A_{p,{\bf j},\sigma}$}

The explicit expression of the transformed Hamiltonian in (\ref{Keq4}), hence also the
explicit form of the $\hat A_{p,{\bf j},\sigma}$ operators, can be find by solving the
matching conditions (\ref{Keq5},\ref{Keq6}). These are representing 36 coupled nonlinear
complex algebraic equations containing 33 unknown variables, namely the numerical
prefactors $a_i,b_i,d_i,...,k_i$ in (\ref{Keq3}), (in total 32 variables), and the 
constant $K$ 
in (\ref{Keq4}). 
Based on the solution deduced in the Appendix A, see (\ref{EquA13}), the introduced
block operators have the expression
\begin{eqnarray}
\hat A_{1,{\bf j},\sigma} &=& \sqrt{t_4} \: [ \:
(\hat c_{{\bf j}+{\bf r}_2,\sigma} + \hat c_{{\bf j}+{\bf r}_3,\sigma}) + 
\frac{t'_1}{t_4} (\hat c_{{\bf j},\sigma} + \hat c_{{\bf j}+{\bf r}_4,\sigma}) \: ],
\nonumber\\
\hat A_{2,{\bf j},\sigma} &=& \sqrt{t_4} \: [ \:
(\frac{t'_2}{t'_1} - \frac{t_1+2t'_2}{2t_3}) ( \hat c_{{\bf j}+{\bf r}_5,\sigma} - 
\hat c_{{\bf j}+{\bf r}_6,\sigma}) + \frac{t'_1}{t_4} (\hat c_{{\bf j},\sigma} - 
\hat c_{{\bf j}+{\bf r}_4,\sigma}) \: ],  
\nonumber\\
\hat A_{3,{\bf j},\sigma} &=& \sqrt{t_4} \: [ \:
(\frac{t'_2}{t'_1} - \frac{t_1+2t'_2}{2t_3}) ( \hat c_{{\bf j}+{\bf a},\sigma} - 
\hat c_{{\bf j}+{\bf r}_4,\sigma}) + \frac{t'_1}{t_4} 
(\hat c_{{\bf j}+{\bf r}_5,\sigma} - \hat c_{{\bf j}+{\bf r}_6+{\bf a},\sigma}) \: ],
\nonumber\\
\hat A_{4,{\bf j},\sigma} &=& \sqrt{t_4} \: [ \:
(\hat c_{{\bf j}+{\bf r}_7,\sigma} + \hat c_{{\bf j}+{\bf r}_8,\sigma}) +
\frac{t'_1}{t_4} (\hat c_{{\bf j}+{\bf r}_5,\sigma} + 
\hat c_{{\bf j}+{\bf r}_6+{\bf a},\sigma}) \: ],  
\label{Keq7}\\
\hat A_{5,{\bf j},\sigma} &=& t'_1 \sqrt{\frac{t_1+2t'_2}{2t'_1t_3}} \: [ \: 
(\hat c_{{\bf j}+{\bf a},\sigma} + \hat c_{{\bf j}+{\bf r}_5,\sigma}) +
\frac{2t_3}{t_1+2t'_2} \hat c_{{\bf j}+{\bf r}_3,\sigma} + \frac{(t_3-t'_1)}{t'_1}
\hat c_{{\bf j}+{\bf r}_4,\sigma} \: ], 
\nonumber\\
\hat A_{6,{\bf j},\sigma} &=& 
t'_1 \sqrt{\frac{t_1+2t'_2}{2t'_1t_3}} \: [ \:
(\hat c_{{\bf j}+{\bf r}_4,\sigma} + \hat c_{{\bf j}+{\bf r}_6+{\bf a},\sigma})+
\frac{2t_3}{t_1+2t'_2} \hat c_{{\bf j}+{\bf r}_2+{\bf a},\sigma} + 
\frac{(t_3-t'_1)}{t'_1} \hat c_{{\bf j}+{\bf a},\sigma} \: ], 
\nonumber\\
\hat A_{7,{\bf j},\sigma} &=& 
t'_1 \sqrt{\frac{t_1+2t'_2}{2t'_1t_3}} \: [ \:
(\hat c_{{\bf j}+{\bf a},\sigma} + \hat c_{{\bf j}+{\bf r}_5+{\bf a},\sigma}) +
\frac{2t_3}{t_1+2t'_2} \hat c_{{\bf j}+{\bf r}_8,\sigma} + 
\frac{(t_3-t'_1)}{t'_1} \hat c_{{\bf j}+{\bf r}_6+{\bf a},\sigma} \: ], 
\nonumber\\
\hat A_{8,{\bf j},\sigma} &=& 
t'_1 \sqrt{\frac{t_1+2t'_2}{2t'_1t_3}} \: [ \: 
(\hat c_{{\bf j}+{\bf r}_4,\sigma} + \hat c_{{\bf j}+{\bf r}_6,\sigma}) + 
\frac{2t_3}{t_1+2t'_2} \hat c_{{\bf j}+{\bf r}_7,\sigma} + \frac{(t_3-t'_1)}{
t'_1} \hat c_{{\bf j}+{\bf r}_5,\sigma} \: ] . 
\nonumber
\end{eqnarray}
The parameter space domain where this solution of the matching conditions is valid
is not severely restricted, being given by
$t_4 > 0$,  $t_3 t'_1/(t_1+2t'_2) > 0$, and  (\ref{EquA11}), see Appendix A.
The constant $K$ present in the transformed Hamiltonian from (\ref{Keq4}) is given 
in this case by
\begin{eqnarray}
 K=\epsilon_1-t_4 - \frac{2t_3 t'_1}{(t_1+2t'_2)}.
\label{Keq8}
\end{eqnarray}

One notes that the $\hat A_{p,{\bf j},\sigma}$ operators present in (\ref{Keq7}) 
are Fermi operators,
since $\hat A^{\dagger}_{p,{\bf j},\sigma} \hat A^{\dagger}_{p,{\bf j},\sigma} =
\hat A_{p,{\bf j},\sigma} \hat A_{p,{\bf j},\sigma} =0$ is satisfied. But these operators
are not canonical Fermi operators, since $\{ \hat A_{p,{\bf j},\sigma},
 \hat A^{\dagger}_{p',{\bf j}',\sigma'} \} \ne \delta_{p,p'} \delta_{{\bf j},{\bf j}'}
\delta_{\sigma,\sigma'}$ holds. 

\section{The deduction of the ground state wave function}

\subsection{The deduction method}

Based on (\ref{Keq7},\ref{Keq8}) now one knows the explicit expression of the
transformed Hamiltonian (\ref{Keq4}), hence one starts the deduction of the
ground state wave function for it, which is written in the form
\begin{eqnarray}
|\Psi_g\rangle = [\prod_{\alpha \in I} \hat B_{\alpha}^{\dagger} ] |0\rangle.
\label{Keq9}
\end{eqnarray}
In (\ref{Keq9}), $|0\rangle$ represents the bare vacuum, $\hat B^{\dagger}_{\alpha}$ 
are linear
combinations of the starting canonical Fermi creation operators $\hat c^{\dagger}_{
{\bf i},\sigma}$, $\alpha$ represents an index which labels different 
$\hat B^{\dagger}$ operators, and $I$ represents the domain from which the $\alpha$
indices must be choosen in order to build up the ground state wave function.

The strategy in deducing the $\hat B^{\dagger}_{\alpha}$ operators is based on the
positive semidefinite structure of $\hat H$ from (\ref{Keq4}), namely on the fact 
that $\hat H'=\hat H - K \hat N = \hat H_A + \hat H_U$ is a positive semidefinite
operator, where $\hat H_A = \sum_{\sigma} \sum_{p=1}^8 \sum_{{\bf j}=1}^{N_c} 
\hat H_{A,p,{\bf j},\sigma}$ and $\hat H_{A,p,{\bf j},\sigma}= 
\hat A_{p,{\bf j},\sigma}^{\dagger} \hat A_{p,{\bf j},\sigma}$. Consequently, 
$|\Psi_g \rangle$ is the most general wave vector satisfying the requirement
\begin{eqnarray}
(\hat H_A + \hat H_U) |\Psi_g \rangle = 0,
\label{Keq10}
\end{eqnarray}
which, because both $\hat H_A$ and $\hat H_U$ are independently positive semidefinite, 
is satisfied only if both $\hat H_A |\Psi_g\rangle =0$, and $\hat H_U |\Psi_g\rangle = 0$
hold. But since the kernel $ker(\hat O)$, of an operator $\hat O$, is a Hilbert subspace
spanned by all vectors $|\phi\rangle$ satisfying $\hat O |\phi\rangle = 0$, it results 
that
in order to deduce $|\Psi_g\rangle$ we must deduce the most general wave vectors placed
inside 
\begin{eqnarray}
ker(\hat H')= ker(\hat H_A) \cap ker(\hat H_U).
\label{Keq10a}
\end{eqnarray}   

Concentrating first on $ker(\hat H_A)$, one observes that since 
$\hat H_{A,p,{\bf j},\sigma}$
are all positive semidefinite operators for all values of all indices, it results that
\begin{eqnarray}
&&ker(\hat H_A)=ker(\hat H_{A,p=1,{\bf j}_1,\uparrow}) \cap ker(\hat H_{A,p=1,{\bf j}_1,
\downarrow}) \cap ker(\hat H_{A,p=2,{\bf j}_1,\uparrow}) \cap ker(\hat H_{A,p=2,{\bf j}_1,
\downarrow}) \cap .... 
\nonumber\\
&&\hspace*{2cm}...\cap ker(\hat H_{A,p=8,{\bf j}_{N_c},\uparrow}) \cap 
ker(\hat H_{A,p=8,{\bf j}_{N_c},\downarrow}),
\label{Keq11}
\end{eqnarray}
where, in the right side of the equality, all indices $p,{\bf j},\sigma$ occur. 

Furthermore, one knows that for fixed 
$p=p',{\bf j}={\bf j}',\sigma=\sigma'$ indices, $ker(\hat H_{A,p',{\bf j}',\sigma'})$ is 
spanned by vectors of the form $|\Psi_{A,p',{\bf j}',\sigma'}\rangle = \prod_{\alpha \in 
I_{A,p',{\bf j}',\sigma'}} \hat B^{\dagger}_{\alpha} |0\rangle$ where 
$I_{A,p',{\bf j}',\sigma'}$ collects all $\alpha$ indices for which the anti-commutation
relation $\{\hat A_{p',{\bf j}',\sigma'},\hat B_{\alpha}^{\dagger}\} = 0$ holds 
\cite{GKV2}. 
Indeed in this case 
$\hat H_{A,p',{\bf j}',\sigma'} |\Psi_{A,p',{\bf j}',\sigma'}\rangle =
\hat A^{\dagger}_{p',{\bf j}',\sigma'} \hat A_{p',{\bf j}',\sigma'} 
\prod_{\alpha \in I_{A,p',{\bf j}',\sigma'}} \hat B_{\alpha}^{\dagger} |0\rangle=0$
is satisfied,
since based on the prescribed anti-commutation relation $\hat A_{p',{\bf j}',\sigma'} 
\hat B_{\alpha}^{\dagger} |0\rangle= - \hat B_{\alpha}^{\dagger}
\hat A_{p',{\bf j}',\sigma'} |0\rangle=0$, which given by 
$\hat A_{p',{\bf j}',\sigma'} |0\rangle=0$, is indeed a true relation.

It results that based on (\ref{Keq11}), the vectors contained in  $ker(\hat H_A)$ can be 
deduced by determining $\hat B^{\dagger}_{\alpha}$ from the relation

\begin{eqnarray}
\{\hat A_{p,{\bf j},\sigma},\hat B_{\alpha}^{\dagger}\} = 0,
\label{Keq12}
\end{eqnarray}
where (\ref{Keq12}) must be satisfied for all values of all indices. One notes that 
different $\hat B^{\dagger}_{\alpha}$ solutions of (\ref{Keq12}) must be linearly 
independent.

After obtaining the $\hat B^{\dagger}_{\alpha}$ operators from (\ref{Keq12}),
the domain $I$
present in (\ref{Keq9}) is such fixed to push $|\Psi_g\rangle$ also in 
$ker(\hat H_U)$. This
last kernel is simply the Hilbert subspace containing all wave vectors with zero double
occupancy. 

\subsection{The calculation of $\hat B^{\dagger}_{\alpha}$ operators}

\subsubsection{The $\hat B^{\dagger}_{\alpha}$ expression}

In deducing the $\hat B^{\dagger}_{\alpha}$ operators one starts from (\ref{Keq12}) and
the already deduced expressions of the $\hat A_{p,{\bf j},\sigma}$ block operators 
presented in (\ref{Keq7}). One considers
\begin{eqnarray}
\hat B^{\dagger}_{\alpha} = \sum_{{\bf i} \in {\cal{D}}_{B_{\alpha}}} x_{\alpha,{\bf i}} 
\hat c^{\dagger}_{{\bf i},\sigma_{\bf i}},
\label{Keq13}
\end{eqnarray}
where two kind of unknown variables must be determined, namely i) the domain ${\cal{D}}_{
B_{\alpha}}$ on which the operator $\hat B^{\dagger}_{\alpha}$ is defined, and ii) the
numerical prefactors $x_{\alpha,{\bf i}}$ which specify the linear combination on which 
$\hat B^{\dagger}_{\alpha}$ is constructed. After fixing the $\hat B^{\dagger}_{\alpha}$
expression in (\ref{Keq13}), the equation (\ref{Keq12}) must be written for all
$\hat A_{p,{\bf j},\sigma}$ block operators, obtaining a coupled system of $16 N_c$ 
equations
which must be solved.

In performing this job one observes the following properties. a) The domain 
${\cal{D}}_{B_{\alpha}}$ cannot be restricted (e.g. extended over $N_b < N_c$ cells). The
reason for this is that for restricted ${\cal{D}}_{B_{\alpha}}$, we always find at least 
one
$\hat A_{p,{\bf j},\sigma}$ operator (holding some fixed indices $p=p', {\bf j}={\bf j}',
\sigma=\sigma'$) which touches the $\hat B^{\dagger}_{\alpha}$ operator only in a 
single point
${\bf j}^{*}$. In this case, since only one kind of starting fermionic operator is acting
on each site and $\{\hat c_{{\bf j}^*,\sigma}, \hat c^{\dagger}_{{\bf j}^*,\sigma} \}=1$
holds, it results that $\{\hat A_{p',{\bf j}',\sigma'}, \hat B^{\dagger}_{\alpha} \} =0$ 
cannot be satisfied. Consequently, the $\hat B^{\dagger}_{\alpha}$ operators will be 
extended operators (i.e. the domain ${\cal{D}}_{B_{\alpha}}$ will be extended along the
whole system). b) For similar reasons it is not possible to have in the expression of
$\hat B^{\dagger}_{\alpha}$ operators islands of $\hat c^{\dagger}_{{\bf i}',\sigma'}$ 
operators in the background of $\hat c^{\dagger}_{{\bf i},\sigma}$ operators where 
$\sigma' \ne \sigma$. Consequently, for a given $\hat B^{\dagger}_{\alpha}$ in 
(\ref{Keq13}) all $\sigma_{\bf i}$ indices become fixed, hence $\sigma_{\bf i}=\sigma$
holds. 

Given by these properties, the $\alpha$ index in (\ref{Keq13}) specifies in fact a 
fixed spin 
index $\sigma$, and a label index $\ell$ which numbers the possible $\hat B^{\dagger}$
operators, hence $\alpha = (\ell,\sigma)$, and the expression of the $\hat B^{\dagger}$
operators from (\ref{Keq13}) becomes
\begin{eqnarray}
\hat B^{\dagger}_{\ell,\sigma} = \sum_{\bf i} x_{\ell,{\bf i}} 
\hat c^{\dagger}_{{\bf i},\sigma},
\label{Keq14}
\end{eqnarray}
where the ${\bf i}$ sum extends over all sites of the system (in all sublattices).
  
\subsubsection{The calculation procedure}

For the effective calculation of the $x_{\ell,{\bf i}}$ numerical prefactors of the
extended operators as $\hat B^{\dagger}_{\ell,\sigma}$ present in (\ref{Keq14}), 
special techniques
are needed. Such a technique has been developed and presented in Ref.\cite{TKG}, 
based on a recursive connection between coefficients in neighboring cells described
by a non-symmetric matrix $\tilde R$, the prefactors being deduced via the unity 
eigenvalues of ${\tilde R}^m$, $m$ being an arbitrary positive integer. In order to 
enlarge and develop the techniques applicable in treating extended operators, one
presents and uses here another alternative method. This is based on deducing the
$\hat B^{\dagger}_{\ell,\sigma}$ operators for a relatively small chain -- named test 
chain -- containing $m$ cells, the test chain being closed in a ring by periodic boundary 
conditions. Once these operators are deduced, the obtained result is used
in the description of a long chain containing $M = p \times m$ 
cells, the long chain being as well treated by periodic boundary conditions, $p$ being 
an arbitrary positive integer.   

In this method first one deduces the linearly independent 
$\hat B^{\dagger}_{\ell,\sigma}$ operators for the test chain. This last contains 
$m$ cells, which are in order numbered by the index $n_c=1,2,...,m$. The coefficients
$x_{\ell,{\bf i}}$ become in this way $x_{\ell,n_c,\nu}$ where $\nu$ denotes 
the lattice site ${\bf j}+{\bf r}_{\nu}$, $\nu=1,2,...,8$ inside the $n_c$th cell 
defined at the site ${\bf j}$ of the $S_1$ sublattice (see Fig.1). Consequently, in
this notation, by changing the sum over ${\bf i}$ in a sum over $n_c$ and $\nu$, 
(\ref{Keq14}) becomes
\begin{eqnarray}
\hat B^{\dagger}_{\ell,\sigma} = \sum_{n_{c_{\bf j}}=1}^m \sum_{\nu=1}^8 
x_{\ell,n_{c_{\bf j}},\nu} \hat c^{\dagger}_{{\bf j}+ {\bf a} n_{c_{\bf j}} +
{\bf r}_{\nu},\sigma},
\label{Keq15}
\end{eqnarray}   
where ${\bf j}$ is an arbitrary site of the sublattice $S_1$.
Based on (\ref{Keq15}) and (\ref{Keq3}), the anti-commutation relation (\ref{Keq12}) 
provides the following system of equations for the prefactors of the 
$\hat B^{\dagger}_{\ell,\sigma}$ operators

\begin{eqnarray}
&&a_1 x_{\ell,1,1} + a_2 x_{\ell,1,2} + a_3 x_{\ell,1,3} + a_4 x_{\ell,1,4} = 0,
\nonumber\\
&&b_1 x_{\ell,1,1} + b_4 x_{\ell,1,4} + b_5 x_{\ell,1,5} + b_6 x_{\ell,1,6} = 0,
\nonumber\\
&&d_0 x_{\ell,2,1} + d_4 x_{\ell,1,4} + d_5 x_{\ell,1,5} + d_6 x_{\ell,2,6} = 0,
\nonumber\\
&&e_5 x_{\ell,1,5} + e_6 x_{\ell,2,6} + e_7 x_{\ell,1,7} + e_8 x_{\ell,1,8} = 0,
\nonumber\\
&&f_0 x_{\ell,2,1} + f_3 x_{\ell,1,3} + f_4 x_{\ell,1,4} + f_5 x_{\ell,1,5} = 0,
\nonumber\\
&&g_0 x_{\ell,2,1} + g_2 x_{\ell,2,2} + g_4 x_{\ell,1,4} + g_6 x_{\ell,2,6} = 0,
\nonumber\\
&&h_0 x_{\ell,2,1} + h_5 x_{\ell,2,5} + h_6 x_{\ell,2,6} + h_8 x_{\ell,1,8} = 0,
\nonumber\\
&&k_4 x_{\ell,1,4} + k_5 x_{\ell,1,5} + k_6 x_{\ell,1,6} + k_7 x_{\ell,1,7} = 0,
\nonumber\\
&&a_1 x_{\ell,2,1} + a_2 x_{\ell,2,2} + a_3 x_{\ell,2,3} + a_4 x_{\ell,2,4} = 0,
\nonumber\\
&&b_1 x_{\ell,2,1} + b_4 x_{\ell,2,4} + b_5 x_{\ell,2,5} + b_6 x_{\ell,2,6} = 0,
\nonumber\\
&&d_0 x_{\ell,3,1} + d_4 x_{\ell,2,4} + d_5 x_{\ell,2,5} + d_6 x_{\ell,3,6} = 0,
\nonumber\\
&&e_5 x_{\ell,2,5} + e_6 x_{\ell,3,6} + e_7 x_{\ell,2,7} + e_8 x_{\ell,2,8} = 0,
\nonumber\\
&&f_0 x_{\ell,3,1} + f_3 x_{\ell,2,3} + f_4 x_{\ell,2,4} + f_5 x_{\ell,2,5} = 0,
\nonumber\\
&&g_0 x_{\ell,3,1} + g_2 x_{\ell,3,2} + g_4 x_{\ell,2,4} + g_6 x_{\ell,3,6} = 0,
\nonumber\\
&&h_0 x_{\ell,3,1} + h_5 x_{\ell,3,5} + h_6 x_{\ell,3,6} + h_8 x_{\ell,2,8} = 0,
\nonumber\\
&&k_4 x_{\ell,2,4} + k_5 x_{\ell,2,5} + k_6 x_{\ell,2,6} + k_7 x_{\ell,2,7} = 0,
\nonumber\\
&&.................................................................
\nonumber
\end{eqnarray}
\begin{eqnarray}
&&a_1 x_{\ell,m,1} + a_2 x_{\ell,m,2} + a_3 x_{\ell,m,3} + a_4 x_{\ell,m,4} = 0,
\nonumber\\
&&b_1 x_{\ell,m,1} + b_4 x_{\ell,m,4} + b_5 x_{\ell,m,5} + b_6 x_{\ell,m,6} = 0,
\nonumber\\
&&d_0 x_{\ell,1,1} + d_4 x_{\ell,m,4} + d_5 x_{\ell,m,5} + d_6 x_{\ell,1,6} = 0,
\nonumber\\
&&e_5 x_{\ell,m,5} + e_6 x_{\ell,1,6} + e_7 x_{\ell,m,7} + e_8 x_{\ell,m,8} = 0,
\nonumber\\
&&f_0 x_{\ell,1,1} + f_3 x_{\ell,m,3} + f_4 x_{\ell,m,4} + f_5 x_{\ell,m,5} = 0,
\nonumber\\
&&g_0 x_{\ell,1,1} + g_2 x_{\ell,1,2} + g_4 x_{\ell,m,4} + g_6 x_{\ell,1,6} = 0,
\nonumber\\
&&h_0 x_{\ell,1,1} + h_5 x_{\ell,1,5} + h_6 x_{\ell,1,6} + h_8 x_{\ell,m,8} = 0,
\nonumber\\
&&k_4 x_{\ell,m,4} + k_5 x_{\ell,m,5} + k_6 x_{\ell,m,6} + k_7 x_{\ell,m,7} = 0,
\label{Keq16}
\end{eqnarray}
where the first eight equations from (\ref{Keq16}) represent (\ref{Keq12}) written with 
the operators
from (\ref{Keq3}) defined in the first cell, the second eight equations from (\ref{Keq16})
represent 
(\ref{Keq12}) written with the operators from (\ref{Keq3}) defined in the second cell, 
and so on, the last eight equalities from (\ref{Keq16})
represent (\ref{Keq12}) written with the 
operators from (\ref{Keq3}) defined in the last $m$th cell. 
The coefficients $a_i,...,k_i$ are presented in (\ref{EquA12}) or (\ref{Keq7}).
One finds in this manner a system of $8m$ linear homogeneous equations for $8m$ unknown 
variables, the $x_{\ell,n_c,\nu}$ prefactors, $\ell$ being fixed. Solutions are present 
only if the determinant of the system (\ref{Keq16}) is zero. This condition specifies
different interconnection possibilities between $\hat H_0$ parameters (e.g. different
regions of the parameter space specified by the index $\gamma$) in a generic form
\begin{eqnarray}
F_{\gamma}(\{ t_{\alpha}\}, \{ \epsilon_{\alpha} \} ) =0,
\label{Keq17}
\end{eqnarray}
where solutions are possible to occur. In (\ref{Keq17}) the sets $\{ t_{\alpha} \}$ and
$\{\epsilon_{\alpha}\}$ are representing the set of hopping matrix elements, and on-site 
one-particle potentials, respectively. The study of the physical meaning of the condition
(\ref{Keq17}) shows that since flat bands are not present in the bare band structure
(see Appendix B, Eq.(\ref{b7}) together with explications related to this equation), 
(\ref{Keq17}) is not connected to the flat band notion, consequently, the ferromagnetism 
which will be here deduced is not of flat band type. Instead, (\ref{Keq17}) specifies 
regions in the parameter space where the density of states around the minimum system 
energy value is high. 

One given deduced $\hat B^{\dagger}$ operator from (\ref{Keq16}) receives a fixed 
$\ell$ index,
for example $\ell=1$. It has hence the form
\begin{eqnarray}
\hat B^{\dagger}_{1,\sigma} = \sum_{n=1}^m \sum_{\nu=1}^8 
x_{1,n,\nu} \hat c^{\dagger}_{{\bf j}+n{\bf a}+{\bf r}_{\nu},\sigma} \: ,
\label{Keq18}
\end{eqnarray}
where ${\bf j}$ is an arbitrary site of the sublattice $S_1$.
The obtained $x_{1,n,\nu}$ coefficients in (\ref{Keq18}) usually have the following 
properties:
i) are  different in different cells,
e.g. for $n \ne n'$ one has $x_{1,n,\nu} \ne x_{1,n',\nu}$, and ii) rotating the 
$\hat B^{\dagger}_{1,\sigma}$ operator by $\pi$ angle around the axis (the line) 
of the chain, the prefactors attached to each site change their value, e.g.
\begin{eqnarray}
&&x_{1,n,1} \ne x_{1,n-1,5}, \quad x_{1,n,2} \ne x_{1,n-1,7}, \quad x_{1,n,3} \ne x_{1,n-1,8},
\quad x_{1,n,4} \ne x_{1,n,6},
\nonumber\\
&&x_{1,n,5} \ne x_{1,n,1}, \quad x_{1,n,6} \ne x_{1,n-1,4}, \quad
x_{1,n,7} \ne x_{1,n,2}, \quad x_{1,n,8} \ne x_{1,n,3}.
\label{Keq19}
\end{eqnarray}
Given by the properties i), ii), new linearly independent $\hat B^{\dagger}_{\ell,\sigma}$
operators are obtained as follows.

i) One translates all coefficients in (\ref{Keq18}) with one cell,
the procedure being possible to be repeated $m-1$ times. One obtains in this manner
\begin{eqnarray}
&&\hat B^{\dagger}_{2,\sigma} = \sum_{n=1}^m \sum_{\nu=1}^8 
x_{1,n-1,\nu} \hat c^{\dagger}_{{\bf j}+n{\bf a}+{\bf r}_{\nu},\sigma},
\nonumber\\
&&\hat B^{\dagger}_{3,\sigma} = \sum_{n=1}^m \sum_{\nu=1}^8 
x_{1,n-2,\nu} \hat c^{\dagger}_{{\bf j}+n{\bf a}+{\bf r}_{\nu},\sigma},
\nonumber\\
&&...........................
\nonumber\\
&&\hat B^{\dagger}_{m,\sigma} = \sum_{n=1}^m \sum_{\nu=1}^8 
x_{1,n-m+1,\nu} \hat c^{\dagger}_{{\bf j}+n{\bf a}+{\bf r}_{\nu},\sigma}.
\label{Keq20}
\end{eqnarray} 

ii) After this step one rotates by $\pi$ degree around the axis of the chain all operators
$\hat B^{\dagger}_{\ell,\sigma}$, $\ell=1,2,...,m$, obtaining $m$ new linearly independent
$\hat B^{\dagger}_{\ell,\sigma}$ operators for which $\ell=m+1,m+2,...,2m$ holds. 
The expression of a
$\hat B^{\dagger}_{\ell',\sigma}$ operator obtained by rotating 
$\hat B^{\dagger}_{\ell,\sigma}$ from (\ref{Keq15}) becomes
\begin{eqnarray}
\hat B^{\dagger}_{\ell',\sigma} = \sum_{n_{c_{\bf j}}=1}^m \sum_{\nu=1}^8 
x'_{\ell',n_{c_{\bf j}},\nu} \hat c^{\dagger}_{{\bf j}+ {\bf a} n_{c_{\bf j}}+
{\bf r}_{\nu},\sigma},
\label{Keq21}
\end{eqnarray}     
where ${\bf j}$ is an arbitrary site of $S_1$, and one has (see also (\ref{Keq19}))
\begin{eqnarray}
&&x'_{\ell',n,1}=x_{\ell,n-1,5}, \quad x'_{\ell',n,2}=x_{\ell,n-1,7}, \quad
x'_{\ell',n,3}=x_{\ell,n-1,8}, \quad x'_{\ell',n,4}=x_{\ell,n,6},
\nonumber\\
&&x'_{\ell',n,5}=x_{\ell,n,1}, \quad x'_{\ell',n,6}=x_{\ell,n-1,4},\quad 
x'_{\ell',n,7}=x_{\ell,n,2}, \quad x'_{\ell',n,8}=x_{\ell,n,3}.
\label{Keq22}
\end{eqnarray}
The linear independence of the new operators obtained via (\ref{Keq20},\ref{Keq21}) 
must be always checked, and the presented procedure leads [for one given solution of
(\ref{Keq16})] to maximum $2m$ linearly independent $\hat B^{\dagger}_{\ell,\sigma}$ 
operators for a fixed spin index $\sigma$. The described procedure than must be repeated
for all independent solutions of (\ref{Keq16}) connected to the same parameter space 
domain
(\ref{Keq17}). Explicit examples for the $m=6$ case (twelve hexagons connected in 
six cells)
are presented in Appendix C.

The deduced $\hat B^{\dagger}_{\ell,\sigma}$ operators  
on the test chain containing $m$ cells,
can be used for 
the description of a chain containing $M=p \times m$ cells as well, as stated before, $p$
being an arbitrary integer number. For this, a given
$\hat B^{\dagger}_{\ell,\sigma}$ operator (maintaining $\ell$ and $\sigma$ fixed)
is periodically repeated $p$ times to cover completely the 
chain containing $M$ cells (treated with periodic boundary
conditions). In order to increase the number of linearly independent 
$\hat B^{\dagger}$ operators relating the same chain containing $M$ cells, one must 
collect
all solutions describing the same parameter space region (\ref{Keq17}) with different 
$m'$ values such to reobtain the same fixed $M$ via $M=p' \times m'$, where both
$p'$ and $m'$ are integers.

\subsection{The ground state wave function}

\subsubsection{The expression of the ground state}

Collecting together all linearly independent $\hat B^{\dagger}_{\ell,\sigma}$ operators 
deduced in the previous subsection and denoting their number by $N_B$,  one has for
$|\Psi(N_B)\rangle = \prod_{\sigma} \prod_{\ell=1}^{N_B} \hat B^{\dagger}_{\ell,\sigma}
|0\rangle$ the property
\begin{eqnarray}
\hat H_A |\Psi(N_B)\rangle = 0.  
\label{Keq23}
\end{eqnarray}
The equality (\ref{Keq23}) is a direct consequence of (\ref{Keq12}) [see also the 
explications presented just above (\ref{Keq12})].
Since the kernel $ker(\hat H_A)$ is intimately connected to (\ref{Keq12}), it 
results that 
in the studied conditions, the wave vector $|\Psi(N_B)\rangle$ is unique in satisfying 
(\ref{Keq23}). 

The ground state is obtained from $|\Psi(N_B)\rangle$ 
based on (\ref{Keq10a})
by taking from it the contributions 
which are present also in $ker(\hat H_U)$, i.e. does not contain double
occupancy. Since all $\hat B^{\dagger}_{\ell,\sigma}$ operators are extended along a 
confined space region defined by the studied chain, different 
$\hat B^{\dagger}_{\ell,\sigma}$ operators holding different $\ell$ indices intersect 
each other on several sites of the system. Consequently, the double occupancy can be 
avoided
only if one fixes to the same value all spin indices of all 
$\hat B^{\dagger}_{\ell,\sigma}$ 
operators. This property remains also true if one decreases the number of the 
$\hat B^{\dagger}$ operators present in the wave vector. Hence,
the ground state wave function at $N \leq N_B$ becomes
\begin{eqnarray}
|\Psi_g(N \leq N_B)\rangle = \prod_{\ell=1}^{N} \hat B^{\dagger}_{\ell,\sigma} |0\rangle,
\label{Keq24}
\end{eqnarray} 
where $\sigma$ is fixed. The corresponding ground state energy according to (\ref{Keq4}) 
is
$E_g=K N$, where the $K$ constant is given in (\ref{Keq8}). The ground state nature arises
since because of the missing double occupancy one has 
$\hat H_U |\Psi_g(N \leq N_B)\rangle =0$, while
because of (\ref{Keq23}), $\hat H_A |\Psi_g(N \leq N_B)\rangle =0$. Hence indeed
$(\hat H_A + \hat H_U) |\Psi_g(N \leq N_B)\rangle = \hat H'|\Psi_g(N \leq N_B)\rangle =0$
is satisfied.

One further note that the described ferromagnetic ground state emerges in the relatively 
low concentration limit. For example, using the explicit solution presented in Appendix C
in the case of a system with $N_c=36$ cells (containing 72 hexagons), the number of
electrons per cell $N/N_c \leq 0.69$ holds, where $max(N)=N_B$ is satisfied. 

\subsubsection{Physical properties of the ground state}

The deduced ground state (\ref{Keq24}) is a fully polarized ferromagnet. Besides, it is a 
conducting state since the operators building up the ground state wave vector are 
extended,
and $\delta \mu = \mu_{+}-\mu_{-}= (E_g(N+1)-E_g(N))-(E_g(N)-E_g(N-1))= 0$ holds for 
$N < N_B$. All itinerant electrons being spin-polarized, such systems are potentially 
applicable in spintronics devices.

The obtained ferromagnetism is not of flat band type, since as shown in Appendix B, 
flat bands are not present in the system. The spin polarization occurs since 
extended operators confined
in the restricted space region of the chain touch each other, hence must correlate their
spin index in order to reduce the double occupancy. A such type of ferromagnetism is 
created by the common effect of correlations and confinement, and was observed in 
other hexagon chains as well \cite{TKG}. For the appearance of this phase, besides the
confined geometry, the following aspects should be present: a) The operators 
$\hat B^{\dagger}$ building up the ground state wave vector must be extended. Such a 
property is not {\it a priori} a chain characteristics, and several chain examples are 
known where it does not occur \cite{GKV1,GKV2,Hon}. 
Regarding this property one mentions that 
always when a such type of behavior was observed for chain structures not possessing
flat bands, {\it external hoppings} (i.e. hopping matrix elements connecting different 
sites from different neighboring cells -- see $t'_1$ in Fig.2) were present. 
Consequently,
when the hopping terms allow the movement along the whole chain (e.g. cage regions where
the hopping amplitudes are missing along a chain section perpendicular to the line
of the chain are not present), 
this seems to be a sufficient condition preserving the extended nature of
$\hat B^{\dagger}$ operators. b) The density of states at and around the minimum 
one-particle energy should be high. This preserves the presence of the fully saturated
ferromagnetic state at arbitrary repulsive $U$ in the low concentration limit and finite
chain length in the presented case. The mechanism however surely works (probably with 
lower
$U$ bound) in the thermodynamic limit as well, but the effective proof  of this statement
remains a challenging problem for further investigations.

One further notes that the described ferromagnetic phase emerges for a sequence of 
different possible interdependences between hopping amplitudes
[see the different $A=(t_1-4t'_2)/(t_1+ 2 t'_2)$ values presented after Eq.(\ref{c5})].
Since it is known that carbon-based nanoribbons are sensitive to uniaxial or 
shear strains
\cite{st1}, and/or external pressure \cite{st2}, the needed interdependences between
$\hat H_0$ parameters can be achieved for example by strain effects 
(see also Ref.\cite{st3}).

\section{Summary and conclusions}

An armchair hexagon chain (i.e. polyphenanthrene)
as a representative of a thin armchair ribbon has been 
described in exact terms by the use of a technique based on positive semidefinite 
operator properties. The deduced ground state for mesoscopic samples, in the small 
concentration limit was shown to be ferromagnetic and metallic. This appears in 
conditions in which flat 
bands are not present in the bare band structure of the system, the spin polarization 
being provided by a common effect of correlations and confinement. Since increasing
the ribbon width there are not present physical reasons to cancel in exact terms
this behavior, at least for a restricted domain of hopping amplitudes we
expect similar effect to occur for thicker ribbons as well. 

\acknowledgements
For Z.G. financial support provided by the Alexander von Humboldt Foundation.

\appendix

\section{The deduction of the numerical prefactors of the $\hat A_{p,{\bf j},\sigma}$
operators}

One starts by using from (\ref{Keq5}) those 14 equations from which every one 
contains only two unknown
variables. From these, namely
\begin{eqnarray}
&& t_4 = a_3^*a_2, \quad t_4 = e_8^*e_7, \quad t'_1 = a_3^*a_1, \quad t'_1 = f_5^*f_3,
\nonumber\\
&& t'_1 = a_4^*a_2, \quad t'_1 = g_2^*g_6, \quad t'_1 = h_8^*h_0, \quad 
t'_1 = e_5^*e_8,
\nonumber\\
&& t'_1 = k_4^*k_7, \quad t'_1 = e_7^*e_6, \quad t'_1 = f_0^*f_3, \quad
t'_1 = g_2^*g_4,
\nonumber\\
&& t'_1 = k_7^*k_6, \quad t'_1 = h_5^*h_8,
\label{EquA1}
\end{eqnarray}
one finds
\begin{eqnarray}
&&a_1 = \frac{t'_1}{a_3^*}, \quad a_2 = \frac{t_4}{a_3^*}, \quad
a_4 = \frac{{t'_1}^*}{t_4^*}a_3, \quad f_0 = \frac{{t'_1}^*}{f_3^*}, \quad
f_5 = \frac{{t'_1}^*}{f_3^*},
\nonumber\\
&&e_5 = \frac{{t'_1}^*}{e_8^*}, \quad e_6 = \frac{t'_1}{t_4^*}e_8, \quad
e_7 = \frac{t_4}{e_8^*}, \quad g_4 = \frac{t'_1}{g_2^*}, \quad
g_6 = \frac{t'_1}{g_2^*},
\nonumber\\
&&h_0 = \frac{t'_1}{h_8^*}, \quad h_5 = \frac{{t'_1}^*}{h_8^*}, \quad
k_4 = \frac{{t'_1}^*}{k_7^*}, \quad k_6 = \frac{t'_1}{k_7^*}.
\label{EquA2}
\end{eqnarray}
Now one uses from (\ref{Keq5}) the 10 equations containing 
respectively four unknown variables,
namely
\begin{eqnarray}
&& t_3 = a_2^*a_1 + g_2^*g_0, \quad t_3 = a_4^*a_3 + f_4^*f_3, \quad
t_3 = e_7^*e_5 + k_7^*k_5,
\nonumber\\
&& t_3 = e_6^*e_8 + h_6^*h_8, \quad t'_2 = b_1^*b_5 + h_0^*h_5, \quad 
t'_2 = b_6^*b_4 + k_6^*k_4, 
\nonumber\\
&& t'_2 = d_0^*d_5 + f_0^*f_5, \quad t'_2 = d_6^*d_4 + g_6^*g_4, \quad
0 = a_4^*a_1 + b_4^*b_1,
\nonumber\\
&& 0 = d_6^*d_5 + e_6^*e_5,
\label{EquA3}
\end{eqnarray}
in which, (\ref{EquA2}) is introduced, and provides
\begin{eqnarray}
&&b_4 = - \frac{({t'_1}^*)^2}{t_4^*b_1^*}, \quad
b_5 = \frac{1}{b_1^*} \:[t'_2 - \frac{({t'_1}^*)^2}{|h_8|^2}], \quad
b_6 = b_1 \frac{t_4}{{t'_1}^2} \:[\frac{{t'_1}^2}{|k_7|^2} - {t'_2}^*],
\nonumber\\
&&d_0 = \frac{1}{d_5^*} \:[{t'_2}^* - \frac{|t'_1|^2}{|f_3|^2}], \quad
d_4 = d_5 \frac{t_4}{({t'_1}^*)^2} \:[\frac{|t'_1|^2}{|g_2|^2} - t'_2], \quad 
d_6 = - \frac{{t'_1}^2}{t_4^*d_5^*},
\nonumber\\
&&f_4 = \frac{1}{f_3^*} \:[t_3^* - \frac{{t'_1}^*|a_3|^2}{t_4^*}], \quad
g_0 = \frac{1}{g_2^*} \:[t_3 - \frac{t'_1t_4^*}{|a_3|^2}], \quad
h_6 = \frac{1}{h_8^*} \:[t_3^* - \frac{t'_1|e_8|^2}{t_4^*}],
\nonumber\\
&&k_5 = \frac{1}{k_7^*} \:[t_3 - \frac{{t'_1}^*t_4^*}{|e_8|^2}].
\label{EquA4}
\end{eqnarray}
After this step one remains with 9 unknown variables 
($a_3,b_1,d_5,e_8,f_3,g_2,h_8,k_7$, and $K$), and 12 equations of the
following form
\begin{eqnarray}
&& t_1 = b_5^*b_4 + d_5^*d_4 + f_5^*f_4 + k_5^*k_4, \quad
 t_1 = b_1^*b_6 + g_0^*g_6 + h_0^*h_6 + d_0^*d_6,
\nonumber\\
&& t_2 = b_6^*b_5 + h_6^*h_5 + k_6^*k_5, \quad
t_2 = d_0^*d_4 + f_0^*f_4 + g_0^*g_4,
\nonumber\\
&& \epsilon_0 = |a_1|^2 + |b_1|^2 + |d_0|^2 + |f_0|^2 + |g_0|^2 + |h_0|^2 + K,
\nonumber\\
&& \epsilon_0 = |a_4|^2 + |b_4|^2 + |d_4|^2 + |f_4|^2 + |g_4|^2 + |k_4|^2 + K,
\nonumber\\
&& \epsilon_0 = |b_5|^2 + |d_5|^2 + |e_5|^2 + |f_5|^2 + |h_5|^2 + |k_5|^2 + K,
\nonumber\\
&& \epsilon_0 = |b_6|^2 + |d_6|^2 + |e_6|^2 + |g_6|^2 + |h_6|^2 + |k_6|^2 + K,
\nonumber\\
&& \epsilon_1 = |a_2|^2 + |g_2|^2 + K, \quad
\epsilon_1 = |a_3|^2 + |f_3|^2 + K,
\nonumber\\
&& \epsilon_1 = |e_7|^2 + |k_7|^2 + K, \quad
\epsilon_1 = |e_8|^2 + |h_8|^2 + K.
\label{EquA5}
\end{eqnarray}                                 
Using (\ref{EquA2},\ref{EquA4}) in (\ref{EquA5}), introducing the notations
\begin{eqnarray}
&&|a_3|^2 = x, \quad |d_5|^2 = z, \quad |f_3|^2 = w, \quad |g_2|^2 = t,
\nonumber\\
&&|e_8|^2 = v, \quad |b_1|^2 = y, \quad |h_8|^2 = u, \quad |k_7|^2 = s,
\label{Equ1A6}
\end{eqnarray}
and considering real hopping matrix elements,
the remaining system of 12 equations becomes of the form
\begin{eqnarray}
\epsilon_1 &=& \frac{t_4^2}{x} + t + K, \quad
\epsilon_1 = \frac{t_4^2}{v} + s + K,
\nonumber\\
\epsilon_1 &=& x + w + K, \quad
\epsilon_1 = v + u + K,
\nonumber\\
\epsilon_0 &=& \frac{{t'_1}^2}{t} + 
\frac{{t'_1}^2}{s} + 
\frac{{t'_1}^2x}{t_4^2} + 
\frac{{t'_1}^4}{t_4^2y} + 
\frac{1}{w} \:(t_3 - \frac{t'_1x}{t_4})^2 + 
\frac{t_4^2z}{{t'_1}^4} \:(\frac{{t'_1}^2}{t} - t'_2)^2 + K,
\nonumber\\
\epsilon_0 &=& \frac{{t'_1}^2}{s} + 
\frac{{t'_1}^2}{t} + 
\frac{{t'_1}^2v}{t_4^2} + 
\frac{{t'_1}^4}{t_4^2z} + 
\frac{1}{u} \:(t_3 - \frac{t'_1v}{t_4})^2 + 
\frac{t_4^2y}{{t'_1}^4} \:(\frac{{t'_1}^2}{s} - t'_2)^2 + K,
\nonumber\\
\epsilon_0 &=& z + 
\frac{{t'_1}^2}{v} + 
\frac{{t'_1}^2}{w} + 
\frac{{t'_1}^2}{u} + 
\frac{1}{y} \:(t'_2 - \frac{{t'_1}^2}{u})^2 + 
\frac{1}{s} \:(t_3 - \frac{t'_1t_4}{v})^2 + K,
\nonumber\\
\epsilon_0 &=& y + 
\frac{{t'_1}^2}{x} + 
\frac{{t'_1}^2}{u} + 
\frac{{t'_1}^2}{w} + 
\frac{1}{z} \:(t'_2 - \frac{{t'_1}^2}{w})^2 + 
\frac{1}{t} \:(t_3 - \frac{t'_1t_4}{x})^2 + K,
\nonumber\\
t_2 &=& \frac{t_4}{{t'_1}^2} \:(t'_2 - \frac{{t'_1}^2}{w})(\frac{{t'_1}^2}{t} - t'_2) + 
\frac{t'_1}{w} \:(t_3 - \frac{t'_1x}{t_4}) + 
\frac{t'_1}{t} \:(t_3 - \frac{t'_1t_4}{x}),
\nonumber\\
t_2 &=& \frac{t_4}{{t'_1}^2} \:(t'_2 - \frac{{t'_1}^2}{u})(\frac{{t'_1}^2}{s} - t'_2) + 
\frac{t'_1}{u} \:(t_3 - \frac{t'_1v}{t_4}) + 
\frac{t'_1}{s} \:(t_3 -\frac{t'_1t_4}{v}),
\nonumber\\
t_1 &=& \frac{{t'_1}^2}{t_4y} \:(\frac{{t'_1}^2}{u} - t'_2) + 
\frac{t_4z}{{t'_1}^2} \:(\frac{{t'_1}^2}{t} - t'_2) + 
\frac{t'_1}{w} \:(t_3 - \frac{t'_1x}{t_4}) + 
\frac{t'_1}{s} \:(t_3 - \frac{t'_1t_4}{v}),
\nonumber\\
t_1 &=& \frac{{t'_1}^2}{t_4z} \:(\frac{{t'_1}^2}{w} - t'_2) + 
\frac{t_4y}{{t'_1}^2} \:(\frac{{t'_1}^2}{s} - t'_2) + 
\frac{t'_1}{u} \:(t_3 - \frac{t'_1v}{t_4}) + 
\frac{t'_1}{t} \:(t_3 - \frac{t'_1t_4}{x}).
\label{EquA7}
\end{eqnarray}
The study of (\ref{EquA7}) shows that it provides the same equation for pair of
variables. For example if $\eta$ denotes $x$ or $v$, one finds 
the equation
\begin{eqnarray}
t_2 &=& \frac{t_4}{{t'_1}^2}(t'_2 - \frac{{t'_1}^2}{\epsilon_1-K-\eta})
(\frac{{t'_1}^2}{\epsilon_1-K-\frac{t_4^2}{\eta}}- t'_2)
\nonumber\\ 
&+& \frac{t'_1}{\epsilon_1-K-\eta}(t_3 - \frac{t'_1\eta}{t_4})
+ \frac{t'_1}{\epsilon_1-K-\frac{t_4^2}{\eta}}(t_3 - \frac{t'_1t_4}{\eta}).
\label{EquA8}
\end{eqnarray}
Similar equations are possible to deduce for the $(y,z)$, $(u,w)$, and $(t,s)$
pairs as well, hence one restricts ourselves below to the class of solutions
\begin{eqnarray}
x=v, \quad y=z, \quad u=w, \quad t=s.
\label{EquA9}
\end{eqnarray}
By using (\ref{EquA9}) in (\ref{EquA7}), (\ref{EquA7}) reduces to
six equations of the form
\begin{eqnarray}
\epsilon_1 &=& \frac{t_4^2}{x} + t + K,
\nonumber\\
\epsilon_1 &=& x + w + K,
\nonumber\\
\epsilon_0 &=& \frac{2{t'_1}^2}{t} + 
\frac{{t'_1}^2x}{t_4^2} +
\frac{{t'_1}^4}{t_4^2z} + 
\frac{{t'_1}^2}{w} \:(\frac{t_3}{t'_1} - \frac{x}{t_4})^2 +
\frac{t_4^2z}{{t'_1}^4} \:(\frac{{t'_1}^2}{t} - t'_2)^2 + K,
\nonumber\\
\epsilon_0 &=& z + 
\frac{{t'_1}^2}{x} + 
\frac{2{t'_1}^2}{w} + 
\frac{1}{z} \:(t'_2 - \frac{{t'_1}^2}{w})^2 + 
\frac{{t'_1}^2}{t} \:(\frac{t_3}{t'_1} - \frac{t_4}{x})^2 + K,
\nonumber\\
t_2 &=& -\frac{t_4}{{t'_1}^2} \:(t'_2 - \frac{{t'_1}^2}{w})(t'_2 - \frac{{t'_1}^2}{t}) +
\frac{{t'_1}^2}{w} \:(\frac{t_3}{t'_1} - \frac{x}{t_4}) +
\frac{{t'_1}^2}{t} \:(\frac{t_3}{t'_1} - \frac{t_4}{x}),
\nonumber\\
t_1 &=& -\frac{{t'_1}^2}{t_4z} \:(t'_2 - \frac{{t'_1}^2}{w}) - 
\frac{t_4z}{{t'_1}^2} \:(t'_2 - \frac{{t'_1}^2}{t}) +
\frac{{t'_1}^2}{w} \:(\frac{t_3}{t'_1} - \frac{x}{t_4}) +
\frac{{t'_1}^2}{t} \:(\frac{t_3}{t'_1} - \frac{t_4}{x}).
\label{EquA10}
\end{eqnarray}
One of the possible solutions of (\ref{EquA10}) emerges at $x=v=t_4 > 0$,
$z=y={t'_1}^2/t_4 > 0$, $t=u=s=w= 2t_3 t'_1/(t_1+2t'_2) > 0$,
$K=\epsilon_1-t_4 - 2t_3 t'_1/(t_1+2t'_2)$, and requires
\begin{eqnarray}
t_2 &=& \frac{t_1+2t'_2}{t_3}(t_3-t'_1) - 
\frac{t_4}{{t'_1}^2}\: (t'_2 - t'_1\frac{t_1+2t'_2}{2t_3})^2,
\nonumber\\
\epsilon_0 - \epsilon_1 &=& t'_1 \:[\frac{2t'_1}{t_4} - \frac{2t_3}{t_1+2t'_2}] 
+ t_4 \:[(\frac{t_1+2t'_2}{2t_3})^2 + \frac{{t'_2}^2}{{t'_1}^2} - 1]
\nonumber\\ 
&+& t'_1\frac{t_1+2t'_2}{2t_3} \:[(\frac{t_3}{t'_1}-1)^2 - 
\frac{2t_4t'_2}{{t'_1}^2} + 2].
\label{EquA11}
\end{eqnarray}
It has the form
\begin{eqnarray}
a_1 &=& \frac{t'_1}{\sqrt{t_4}}, \:\:\:\:\: 
a_2 = \sqrt{t_4}, \:\:\:\:\:      
a_3 = \sqrt{t_4}, \:\:\:\:\:
a_4 = \frac{t'_1}{\sqrt{t_4}},                                   
\nonumber\\
b_1 &=& \frac{t'_1}{\sqrt{t_4}}, \:\:\:\:\: 
b_4 = -\frac{t'_1}{\sqrt{t_4}}, \:\:\:\:\:
b_5 = \sqrt{t_4}\: (\frac{t'_2}{t'_1} - \frac{t_1+2t'_2}{2t_3}), \:\:\:\:\:
b_6 = -\sqrt{t_4}\: (\frac{t'_2}{t'_1} - \frac{t_1+2t'_2}{2t_3}),
\nonumber\\
d_0 &=& \sqrt{t_4}\: (\frac{t'_2}{t'_1} - \frac{t_1+2t'_2}{2t_3}), \:\:\:\:\:
d_4 = -\sqrt{t_4}\: (\frac{t'_2}{t'_1} - \frac{t_1+2t'_2}{2t_3}), \:\:\:\:\:
d_5 = \frac{t'_1}{\sqrt{t_4}}, \:\:\:\:\:
d_6 = -\frac{t'_1}{\sqrt{t_4}},                             
\nonumber\\
e_5 &=& \frac{t'_1}{\sqrt{t_4}}, \:\:\:\:\:
e_6 = \frac{t'_1}{\sqrt{t_4}}, \:\:\:\:\:
e_7 = \sqrt{t_4}, \:\:\:\:\:
e_8 = \sqrt{t_4},
\label{EquA12}
\\
f_0 &=& t'_1 \sqrt{\frac{t_1+2t'_2}{2t'_1t_3}}, \:\:\:\:\:
f_3 = \sqrt{\frac{2t'_1t_3}{t_1+2t'_2}}, \:\:\:\:\:
f_4 = (t_3-t'_1)\sqrt{\frac{t_1+2t'_2}{2t'_1t_3}}, \:\:\:\:\:
f_5 = t'_1 \sqrt{\frac{t_1+2t'_2}{2t'_1t_3}},
\nonumber\\
g_0 &=& (t_3-t'_1)\sqrt{\frac{t_1+2t'_2}{2t'_1t_3}}, \:\:\:\:\:
g_2 = \sqrt{\frac{2t'_1t_3}{t_1+2t'_2}}, \:\:\:\:\:
g_4 = t'_1 \sqrt{\frac{t_1+2t'_2}{2t'_1t_3}}, \:\:\:\:\:
g_6 = t'_1 \sqrt{\frac{t_1+2t'_2}{2t'_1t_3}},
\nonumber\\
h_0 &=& t'_1 \sqrt{\frac{t_1+2t'_2}{2t'_1t_3}}, \:\:\:\:\:
h_5 = t'_1 \sqrt{\frac{t_1+2t'_2}{2t'_1t_3}}, \:\:\:\:\:
h_6 = (t_3-t'_1)\sqrt{\frac{t_1+2t'_2}{2t'_1t_3}}, \:\:\:\:\:
h_8 = \sqrt{\frac{2t'_1t_3}{t_1+2t'_2}},
\nonumber\\
k_4 &=& t'_1 \sqrt{\frac{t_1+2t'_2}{2t'_1t_3}}, \:\:\:\:\:
k_5 = (t_3-t'_1)\sqrt{\frac{t_1+2t'_2}{2t'_1t_3}}, \:\:\:\:\:
k_6 = t'_1 \sqrt{\frac{t_1+2t'_2}{2t'_1t_3}}, \:\:\:\:\:
k_7 = \sqrt{\frac{2t'_1t_3}{t_1+2t'_2}}.
\nonumber
\end{eqnarray}
The parameter space region ${\cal{D}}$ where this solution emerges is fixed by 
Eq.(\ref{EquA11}) and the conditions $t_4 > 0$, $t_3 t'_1/(t_1+2t'_2) > 0$. 
The corresponding $\hat A_{p,{\bf j},\sigma}$ operators are provided by the 
following relations
\begin{eqnarray}
\hat A_{1,{\bf j},\sigma} &=& 
\frac{t'_1}{\sqrt{t_4}} \hat c_{{\bf j},\sigma} + 
\sqrt{t_4} \hat c_{{\bf j}+{\bf r}_2,\sigma} + 
\sqrt{t_4} \hat c_{{\bf j}+{\bf r}_3,\sigma} + 
\frac{t'_1}{\sqrt{t_4}} \hat c_{{\bf j}+{\bf r}_4,\sigma},
\nonumber\\
\hat A_{2,{\bf j},\sigma} &=& 
\frac{t'_1}{\sqrt{t_4}} \hat c_{{\bf j},\sigma} - 
\frac{t'_1}{\sqrt{t_4}} \hat c_{{\bf j}+{\bf r}_4,\sigma} + 
\sqrt{t_4}\: (\frac{t'_2}{t'_1} - \frac{t_1+2t'_2}{2t_3}) \hat c_{{\bf j}+{\bf r}_5,\sigma} - 
\sqrt{t_4}\: (\frac{t'_2}{t'_1} - \frac{t_1+2t'_2}{2t_3}) \hat c_{{\bf j}+{\bf r}_6,\sigma},
\nonumber\\
\hat A_{3,{\bf j},\sigma} &=& 
\sqrt{t_4}\: (\frac{t'_2}{t'_1} - \frac{t_1+2t'_2}{2t_3}) \hat c_{{\bf j}+{\bf a},
\sigma} - 
\sqrt{t_4}\: (\frac{t'_2}{t'_1} - \frac{t_1+2t'_2}{2t_3}) \hat c_{{\bf j}+{\bf r}_4,\sigma} + 
\frac{t'_1}{\sqrt{t_4}} \hat c_{{\bf j}+{\bf r}_5,\sigma} - 
\frac{t'_1}{\sqrt{t_4}} \hat c_{{\bf j}+{\bf r}_6+{\bf a},\sigma},
\nonumber\\
\hat A_{4,{\bf j},\sigma} &=& 
\frac{t'_1}{\sqrt{t_4}} \hat c_{{\bf j}+{\bf r}_5,\sigma} + 
\frac{t'_1}{\sqrt{t_4}} \hat c_{{\bf j}+{\bf r}_6+{\bf a},\sigma} + 
\sqrt{t_4} \hat c_{{\bf j}+{\bf r}_7,\sigma} + 
\sqrt{t_4} \hat c_{{\bf j}+{\bf r}_8,\sigma},
\label{EquA13}\\
\hat A_{5,{\bf j},\sigma} &=& 
t'_1 \sqrt{\frac{t_1+2t'_2}{2t'_1t_3}} \hat c_{{\bf j}+{\bf a},\sigma} + 
\sqrt{\frac{2t'_1t_3}{t_1+2t'_2}} \hat c_{{\bf j}+{\bf r}_3,\sigma} + 
(t_3-t'_1)\sqrt{\frac{t_1+2t'_2}{2t'_1t_3}} \hat c_{{\bf j}+{\bf r}_4,\sigma} + 
t'_1 \sqrt{\frac{t_1+2t'_2}{2t'_1t_3}} \hat c_{{\bf j}+{\bf r}_5,\sigma},
\nonumber\\
\hat A_{6,{\bf j},\sigma} &=& 
(t_3-t'_1)\sqrt{\frac{t_1+2t'_2}{2t'_1t_3}} \hat c_{{\bf j}+{\bf a},\sigma} + 
\sqrt{\frac{2t'_1t_3}{t_1+2t'_2}} \hat c_{{\bf j}+{\bf r}_2+{\bf a},\sigma} + 
t'_1 \sqrt{\frac{t_1+2t'_2}{2t'_1t_3}} \hat c_{{\bf j}+{\bf r}_4,\sigma} + 
t'_1 \sqrt{\frac{t_1+2t'_2}{2t'_1t_3}} \hat c_{{\bf j}+{\bf r}_6+{\bf a},\sigma},
\nonumber\\
\hat A_{7,{\bf j},\sigma} &=& 
t'_1 \sqrt{\frac{t_1+2t'_2}{2t'_1t_3}} \hat c_{{\bf j}+{\bf a},\sigma} + 
t'_1 \sqrt{\frac{t_1+2t'_2}{2t'_1t_3}} \hat c_{{\bf j}+{\bf r}_5+{\bf a},\sigma} + 
(t_3-t'_1)\sqrt{\frac{t_1+2t'_2}{2t'_1t_3}} \hat c_{{\bf j}+{\bf r}_6+{\bf a},
\sigma} + 
\sqrt{\frac{2t'_1t_3}{t_1+2t'_2}} \hat c_{{\bf j}+{\bf r}_8,\sigma},
\nonumber\\
\hat A_{8,{\bf j},\sigma} &=& 
t'_1 \sqrt{\frac{t_1+2t'_2}{2t'_1t_3}} \hat c_{{\bf j}+{\bf r}_4,\sigma} + 
(t_3-t'_1)\sqrt{\frac{t_1+2t'_2}{2t'_1t_3}} \hat c_{{\bf j}+{\bf r}_5,\sigma} + 
t'_1 \sqrt{\frac{t_1+2t'_2}{2t'_1t_3}} \hat c_{{\bf j}+{\bf r}_6,\sigma} + 
\sqrt{\frac{2t'_1t_3}{t_1+2t'_2}} \hat c_{{\bf j}+{\bf r}_7,\sigma}.
\nonumber
\end{eqnarray}

\section{The bare band structure}

In order to obtain the bare band structure of the system, one Fourier
transforms $\hat H_0=\hat T_0 + \hat T_1 + \hat T_2$ to ${\bf k}$ space by using 
$\hat c_{{\bf j}+{\bf r}_{\nu}, \sigma} = (1/\sqrt{N_c}) \sum_{\bf k} 
\exp[-i {\bf k}({\bf j}+{\bf r}_{\nu})] \hat c_{\nu,{\bf k},\sigma}$ (where 
the sum over ${\bf k}$ represents a sum over $N_c$ cells), obtaining

\begin{eqnarray}
&&\hat T_0 = \sum_{\sigma} \sum_{\bf k} [ \epsilon_0 (\hat c^{\dagger}_{1,
{\bf k},\sigma}
\hat c_{1,{\bf k},\sigma} + \hat c^{\dagger}_{4,{\bf k},\sigma} \hat c_{4,
{\bf k},\sigma} +
\hat c^{\dagger}_{5,{\bf k},\sigma} \hat c_{5,{\bf k},\sigma} +
\hat c^{\dagger}_{6,{\bf k},\sigma} \hat c_{6,{\bf k},\sigma}) 
\nonumber\\
&&\hspace*{1cm}+ \epsilon_1 (
\hat c^{\dagger}_{2,{\bf k},\sigma} \hat c_{2,{\bf k},\sigma} +
\hat c^{\dagger}_{3,{\bf k},\sigma} \hat c_{3,{\bf k},\sigma} +
\hat c^{\dagger}_{7,{\bf k},\sigma} \hat c_{7,{\bf k},\sigma} +
\hat c^{\dagger}_{8,{\bf k},\sigma} \hat c_{8,{\bf k},\sigma}) \: ],
\nonumber\\
&&\hat T_1 = \sum_{\sigma} \sum_{\bf k} [ t_1 (
\hat c^{\dagger}_{1,{\bf k},\sigma} \hat c_{6,{\bf k},\sigma} e^{-i {\bf k}{\bf r}_6} +
\hat c^{\dagger}_{5,{\bf k},\sigma} \hat c_{4,{\bf k},\sigma} e^{+i {\bf k}({\bf r}_5-{\bf r}_4)} )
\nonumber\\
&&\hspace*{1cm}+ t_2 
(\hat c^{\dagger}_{6,{\bf k},\sigma} \hat c_{5,{\bf k},\sigma} e^{+i {\bf k}({\bf r}_6-{\bf r}_5)} +
\hat c^{\dagger}_{1,{\bf k},\sigma} \hat c_{4,{\bf k},\sigma} e^{+i {\bf k}({\bf a}-{\bf r}_4)})
\nonumber\\
&&\hspace*{1cm}+ t_3 
(\hat c^{\dagger}_{2,{\bf k},\sigma} \hat c_{1,{\bf k},\sigma} e^{+i {\bf k}{\bf r}_2} +
\hat c^{\dagger}_{4,{\bf k},\sigma} \hat c_{3,{\bf k},\sigma} e^{+i {\bf k}({\bf r}_4-{\bf r}_3)}+
\hat c^{\dagger}_{7,{\bf k},\sigma} \hat c_{5,{\bf k},\sigma} e^{+i {\bf k}({\bf r}_7-{\bf r}_5)}+
\hat c^{\dagger}_{6,{\bf k},\sigma} \hat c_{8,{\bf k},\sigma} e^{+i {\bf k}({\bf r}_6+{\bf a}
-{\bf r}_8)}) 
\nonumber\\
&&\hspace*{1cm}+t_4 (
\hat c^{\dagger}_{3,{\bf k},\sigma} \hat c_{2,{\bf k},\sigma} e^{+i {\bf k}({\bf r}_3-{\bf r}_2)}+
\hat c^{\dagger}_{8,{\bf k},\sigma} \hat c_{7,{\bf k},\sigma} e^{+i {\bf k}({\bf r}_8-{\bf r}_7)})
+ H.c. ],
\nonumber\\
&&\hat T_2 = \sum_{\sigma} \sum_{\bf k} [ t'_1 (
\hat c^{\dagger}_{3,{\bf k},\sigma} \hat c_{1,{\bf k},\sigma} e^{+i {\bf k}{\bf r}_3}+
\hat c^{\dagger}_{4,{\bf k},\sigma} \hat c_{2,{\bf k},\sigma} e^{+i {\bf k}({\bf r}_4-{\bf r}_2)}+
\hat c^{\dagger}_{5,{\bf k},\sigma} \hat c_{3,{\bf k},\sigma} e^{+i {\bf k}({\bf r}_5-{\bf r}_3)}
\nonumber\\
&&\hspace*{1cm}+
\hat c^{\dagger}_{2,{\bf k},\sigma} \hat c_{6,{\bf k},\sigma} e^{+i {\bf k}({\bf r}_2-{\bf r}_6)}+
\hat c^{\dagger}_{4,{\bf k},\sigma} \hat c_{7,{\bf k},\sigma} e^{+i {\bf k}({\bf r}_4-{\bf r}_7)}+
\hat c^{\dagger}_{5,{\bf k},\sigma} \hat c_{8,{\bf k},\sigma} e^{+i {\bf k}({\bf r}_5-{\bf r}_8)}
\nonumber\\
&&\hspace*{1cm}+
\hat c^{\dagger}_{7,{\bf k},\sigma} \hat c_{6,{\bf k},\sigma} e^{+i {\bf k}({\bf r}_7-{\bf r}_6-
{\bf a})}+
\hat c^{\dagger}_{8,{\bf k},\sigma} \hat c_{1,{\bf k},\sigma} e^{+i {\bf k}({\bf r}_8-{\bf a})}+
\hat c^{\dagger}_{1,{\bf k},\sigma} \hat c_{3,{\bf k},\sigma} e^{+i {\bf k}({\bf a}-{\bf r}_3)}
\nonumber\\
&&\hspace*{1cm}+
\hat c^{\dagger}_{2,{\bf k},\sigma} \hat c_{4,{\bf k},\sigma} e^{+i {\bf k}({\bf r}_2+{\bf a}
-{\bf r}_4)}+
\hat c^{\dagger}_{7,{\bf k},\sigma} \hat c_{6,{\bf k},\sigma} e^{+i {\bf k}({\bf r}_7-{\bf r}_6)}+
\hat c^{\dagger}_{5,{\bf k},\sigma} \hat c_{8,{\bf k},\sigma} e^{+i {\bf k}({\bf r}_5+{\bf a}
-{\bf r}_8)}) 
\nonumber\\
&&\hspace*{1cm}+ t'_2 (
\hat c^{\dagger}_{1,{\bf k},\sigma} \hat c_{5,{\bf k},\sigma} e^{-i {\bf k}{\bf r}_5} +
\hat c^{\dagger}_{6,{\bf k},\sigma} \hat c_{4,{\bf k},\sigma} e^{+i {\bf k}({\bf r}_6-{\bf r}_4)}+
\hat c^{\dagger}_{1,{\bf k},\sigma} \hat c_{5,{\bf k},\sigma} e^{+i {\bf k}({\bf a}-{\bf r}_5)}
\nonumber\\
&&\hspace*{1cm}+
\hat c^{\dagger}_{6,{\bf k},\sigma} \hat c_{4,{\bf k},\sigma} e^{+i {\bf k}({\bf r}_6+{\bf a}
-{\bf r}_4)}) + H.c. ] .
\label{b1}
\end{eqnarray}
Introducing the notation $ka={\bf k}{\bf a} \in (-\pi,+\pi]$, $b=a/3$, where $a=|{\bf a}|$ 
is the lattice constant, and taking into account that
\begin{eqnarray}
&&{\bf k}{\bf r}_1 =0, \quad {\bf k}{\bf r}_2= \frac{kb}{2}, \quad
{\bf k}{\bf r}_3=\frac{3kb}{2}, \quad {\bf k}{\bf r}_4=2kb,
\nonumber\\
&&{\bf k}{\bf r}_5=\frac{3kb}{2}, \quad {\bf k}{\bf r}_6=\frac{kb}{2}, \quad
{\bf k}{\bf r}_7=2kb, \quad {\bf k}{\bf r}_8= {\bf k}{\bf a}=3kb,
\label{b2}
\end{eqnarray} 
$\hat H_0$ in ${\bf k}$ space becomes of the form
\begin{eqnarray}
\hat H_0 = \sum_{\sigma} \sum_{\bf k} \hat {\tilde C}^{\dagger}_{{\bf k},\sigma}
\tilde M_{\bf k} \hat {\tilde C}_{{\bf k},\sigma},
\label{b3}
\end{eqnarray}
where $\hat {\tilde C}^{\dagger}_{{\bf k},\sigma}$ represents the eight component 
row vector 
$(\hat c^{\dagger}_{1,{\bf k},\sigma}, \hat c^{\dagger}_{2,{\bf k},\sigma}, ...,
\hat c^{\dagger}_{8,{\bf k},\sigma})$, $\hat {\tilde C}_{{\bf k},\sigma}$ 
represents the eight
component column vector obtained as the conjugate transpose of 
$\hat {\tilde C}^{\dagger}_{
{\bf k},\sigma}$, and $\tilde M_{\bf k}$ is 
a $8\times 8$ Hermitian matrix whose transpose $\tilde M^t_{\bf k}$ is
given by

%%%%%%%%%%%%%%%%%%%%%%%%%%%%%%%%%%%%%%%%%%%%%%%%%%%%%%%%%%%%%%%%%%%%%%%%%%
%% FIGURE B1 (Fig.11)
%%%%%%%%%%%%%%%%%%%%%%%%%%%%%%%%%%%%%%%%%%%%%%%%%%%%%%%%%%%%%%%%%%%%%%%%%%
\begin{figure}
%\centerline{\includegraphics[width=10 cm,height=5 cm]{zzF11.eps}}
\centerline{\epsfbox{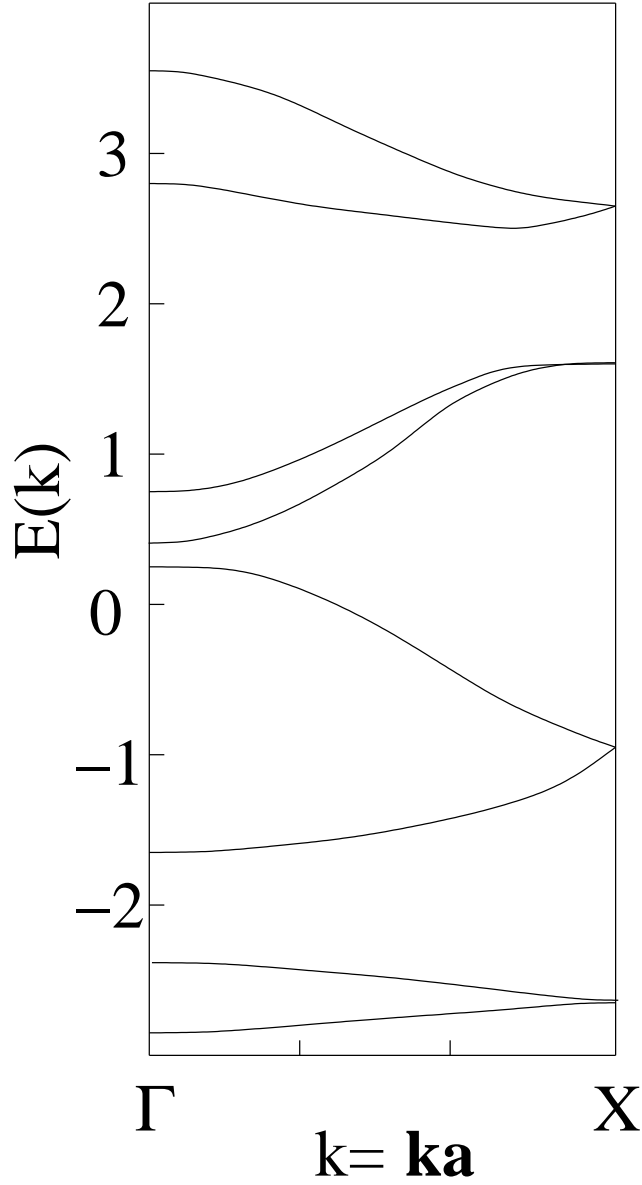}}
\caption{Exemplification of the bare band structure at $t_2/t_1=1.2, t_3/t_1=1.3,
t_4/t_1=1.05, t'_1/t_1=0.1, t'_2/t_1=0.2, \epsilon_0/t_1=-0.9, \epsilon_1/t_1=1.15$,
and $t_1 > 0$. The energy values are in $t_1$ units. Since $E(k={\bf k}{\bf a})$ is 
even in $k$, only the $k\in [0,\pi]$ is shown. The folding at $k=\pi$ is a 
consequence of the structure of the cell.} 
\label{Kfigb1}
\end{figure}
%%%%%%%%%%%%%%%%%%%%%%%%%%%%%%%%%%%%%%%%%%%%%%%%%%%%%%%%%%%%%%%%%%%%%%%%%%

%%%%%%%%%%%%%%%%%%%%%%%%%%%%%%%%%%%%%%%%%%%%%%%%%%%%%%%%%%%%%%%%%%%%%%%%%%
%% FIGURE B2 (Fig12)
%%%%%%%%%%%%%%%%%%%%%%%%%%%%%%%%%%%%%%%%%%%%%%%%%%%%%%%%%%%%%%%%%%%%%%%%%%
\begin{figure}
%\centerline{\includegraphics[width=10 cm,height=5 cm]{zzF12.eps}}
\centerline{\epsfbox{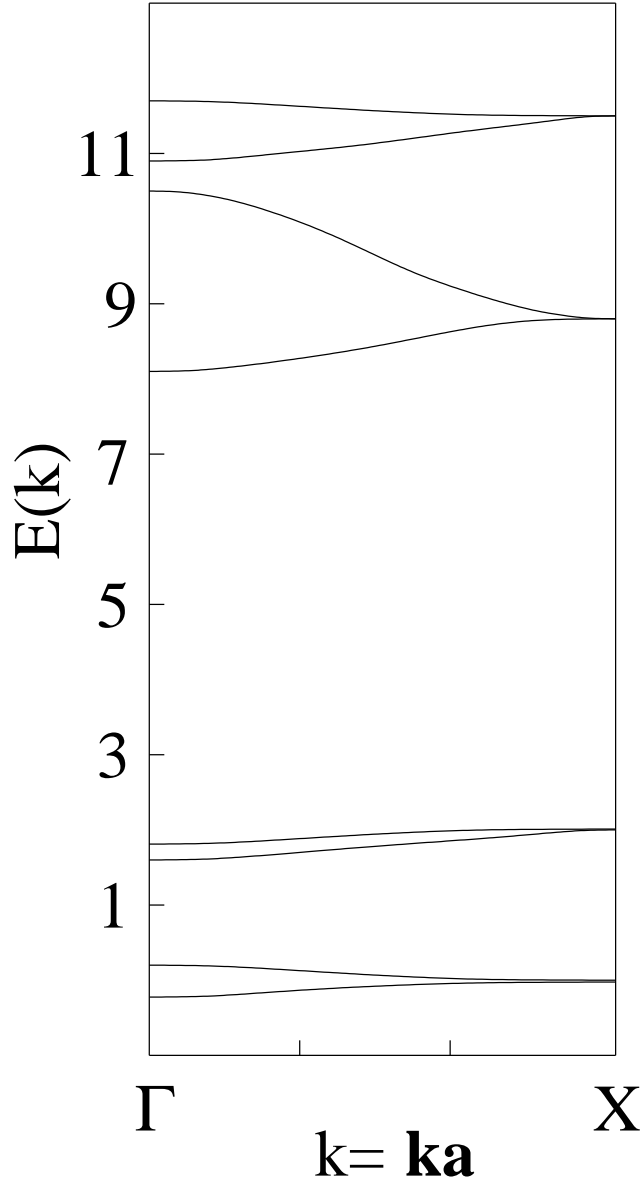}}
\caption{The bare band structure in the presence of the condition
(\ref{EquA11}). One has $t_3/t_1=1.3,
t_4/t_1=1.05, t'_1/t_1=0.1, t'_2/t_1=0.2, \epsilon_1/t_1=1.15$,
$t_1 > 0$, and $t_2,\epsilon_0$ are given by (\ref{EquA11}). 
The energy values are in $t_1$ units. Since $E(k={\bf k}{\bf a})$ is 
even in $k$, only the $k\in [0,\pi]$ is shown. The folding at $k=\pi$ is a 
consequence of the structure of the cell.} 
\label{Kfigb2}
\end{figure}
%%%%%%%%%%%%%%%%%%%%%%%%%%%%%%%%%%%%%%%%%%%%%%%%%%%%%%%%%%%%%%%%%%%%%%%%%%

\begin{eqnarray}
\tilde M^t_{\bf k} = 
\left( \begin{array}{cccccccc}
\epsilon_0 & t_3e^{ikb/2} & 2t'_1 \cos \frac{ak}{2} & t_2e^{-ikb} & 2t'_2 \cos \frac{ak}{2} &
t_1 e^{ikb/2} & 0 & t'_1 \\
t_3 e^{-ikb/2} & \epsilon_1  & t_4 e^{ikb} & 2t'_1 \cos \frac{ak}{2} & 0 & t'_1 & 0 & 0 \\
2 t'_1 \cos \frac{ak}{2} & t_4 e^{-ikb} & \epsilon_1 & t_3 e^{ikb/2} & t'_1 & 0 & 0 & 0  \\
t_2 e^{ikb} & 2 t'_1 \cos \frac{ak}{2} & t_3 e^{-ikb/2} & \epsilon_0 & t_1 e^{-ikb/2} &
2 t'_2 \cos \frac{ak}{2} & t'_1 & 0 \\
2t'_2\cos \frac{ak}{2} & 0 & t'_1 & t_1 e^{ikb/2} & \epsilon_0 & t_2 e^{-ikb} & t_3 e^{ikb/2}
& 2 t'_1 \cos \frac{ak}{2} \\
t_1 e^{-ikb/2} & t'_1 & 0 & 2 t'_2 \cos \frac{ak}{2} & t_2 e^{ikb} & \epsilon_0 & 
2 t'_1 \cos \frac{ak}{2} & t_3 e^{-ikb/2} \\
0 & 0 & 0 & t'_1 & t_3 e^{-ikb/2} & 2 t'_1 \cos \frac{ak}{2} & \epsilon_1 & t_4 e^{ikb} \\
t'_1 & 0 & 0 & 0 & 2 t'_1 \cos \frac{ak}{2} & t_3 e^{ikb/2} & t_4 e^{-ikb} & \epsilon_1 \\
\end{array} \right) .
\label{b4}
\end{eqnarray}
The bare band structure is obtained from the secular equation of $\tilde M_{\bf k}$ as 
an eight 
order algebraic equation in $\lambda = E_n(k)$, $n=1,2,...,8$ of the form 
\begin{eqnarray}
\frac{1}{(e_1^2-t_4^2)^2} \{ (E^2-Z^2-|W|^2 -|F|^2)^2- Y^2 + 4 Z 
[E Y - Z (|W|^2 +|F|^2)]\} =0,
\label{b5}
\end{eqnarray}
where one has
\begin{eqnarray}
&&|F|^2 = |V|^2 +(|W|^2 +2 X) \cos^2 \frac{ak}{2}, \quad Y=2(|W|^2 + X) 
\cos \frac{ak}{2},
\nonumber\\
&&E=e_0(e_1^2-t_4^2) - e_1({t'_1}^2+t_3^2) - 4 t'_1(t'_1e_1-t_3t_4) \cos^2 
\frac{ak}{2},
\nonumber\\
&&Z=2 [ t'_2(e_1^2-t_4^2) - 2 {t'_1}^2 e_1 + t_3 t_4 t'_1] \cos \frac{ak}{2}, 
\quad
X= {\bar d}({\bar a}-{\bar b}) + 4 t_4 {t'_1}^2 ({\bar a}-{\bar b} \cos^2 \frac{ak}{2}),
\nonumber\\
&&|W|^2={\bar d}^2+8t_4{t'_1}^2 {\bar c} \cos^2 \frac{ak}{2}, \quad
|V|^2={\bar a}^2 + {\bar b}({\bar b}-2{\bar a}) \cos^2 \frac{ak}{2}, 
\nonumber\\
&&{\bar a}=t_2(e_1^2-t_4^2) + t_4 ({t'_1}^2-t_3^2), \quad {\bar b}=t_1(e_1^2-t_4^2) 
+ 2 t_3(t'_1e_1-t_3 t_4),
\nonumber\\
&&{\bar c}=t_1(e_1^2-t_4^2) - 2 t'_1(t_3e_1-t'_1 t_4), \quad {\bar d}= t_1(e_1^2-t_4^2) 
- 2 t_3 t'_1e_1,
\nonumber\\
&& e_0 = \epsilon_0 - \lambda, \quad e_1=\epsilon_1 - \lambda .
\label{b6}
\end{eqnarray}
An exemplification of the bare band structure is presented for the general case
in Fig.\ref{Kfigb1}, while the bare band structure in the presence of the condition
(\ref{EquA11}) is exemplified in Fig.\ref{Kfigb2}. As 
Figs.(\ref{Kfigb1},\ref{Kfigb2}) show, flat bands in the bare band structure
are not present for non-zero values of all $\hat H_0$ parameters. This fact can be
directly and analytically verified in the following way. The relation (\ref{b5}) 
can be written as
\begin{eqnarray}
A_8 \cos^8\frac{ak}{2} + A_6\cos^6\frac{ak}{2} + A_4\cos^4\frac{ak}{2} +
A_2 \cos^2\frac{ak}{2} + A_0 = 0,
\label{b7}
\end{eqnarray}
where the numerical prefactors $A_{2m}$, $m=0,1,...,4$ are dependent only on 
$\hat H_0$ parameters and the eigenvalue $\lambda$. Since now in (\ref{b7}) 
the entire $k$ dependence is concentrated in the $\cos^{2m}(ka/2)$ terms, it 
results that flat band solutions (i.e. $k$-independent expressions for $\lambda$)
can be obtained only if for a given solution, simultaneously for all $A_{2m}$, 
$A_{2m}=0$ holds. But, since one has $A_8=(16 {t'_1}^4)^2$, this condition cannot 
be satisfied for non-zero value of the hopping matrix elements. Consequently, 
the band structure created by $\hat H_0$ does not contain flat bands at all. One 
further notes that the $A_{2m < 8}$ prefactors can be as well easily calculated
based on (\ref{b5},\ref{b6}), and one finds that the denominator in (\ref{b5})
simplifies in all $A_{2m}$. One obtains for example
$A_0= \{ (e_1^2-t_4^2)(e_0^2-t_1^2-t_2^2)+({t'_1}^2-t_3^2)^2 +2 [2 t_1 t_3 t'_1
e_1- e_0 e_1 ({t'_1}^2+t_3^2) - t_2 t_4 ({t'_1}^2-t_3^2)] \}^2$, or
$A_6=-256 {t'_1}^2 \{ {t'_1}^2 {t'_2}^2 e_1^2/2 - [t'_2t_4(2t'_2t_3-t'_1t_1)+{t'_1}^2
(2t_1t'_2+t_2t_3)]e_1 t'_1 -{t'_1}^3 t_3t_4e_0 +{t'_1}^4 e_0 e_1 +{t'_1}^4({t'_1}^2 +
t_3^2) +t'_2 t_3 t_4^2 (t'_2t_3-t'_1t_1) +({t'_1}^2t_4/4)[t_4(t_1^2+2{t'_2}^2)+ 2t_3
(4t'_1t'_2+2t_1t'_2+t_2t_3)-4{t'_1}^2(t_1-t_2)] \}$, etc.   

\section{Example for $\hat B^{\dagger}$ operators}

In this section one exemplifies solutions of (\ref{Keq16}) in the $m=6$ case
(i.e. the number of cells in the test chain is six) inside the parameter space region 
specified by
\begin{eqnarray}
t'_1 > 0, \quad t_3 = 3t'_1, \quad 6{t'_1}^2 = t_4(t_1-4t'_2).
\label{c1}
\end{eqnarray}
One further introduces the parameter 
$A=(t_1-4t'_2)/(t_1+ 2 t'_2)$.

In this region one has 
a $\hat B^{\dagger}$ solution holding the same 
coefficients in each cell, the solution being present at arbitrary $N_c=m$ and 
arbitrary $A$. 
This operator will be denoted by the $\ell=1$ index, and has the expression
\begin{eqnarray}
\hat B^{\dagger}_{1,\sigma}= \sum_{n=0}^{N_c-1} (
\hat c^{\dagger}_{{\bf j}+n{\bf a}+{\bf r}_5,\sigma}
- \hat c^{\dagger}_{{\bf j}+n{\bf a}+{\bf r}_4,\sigma}
+ \hat c^{\dagger}_{{\bf j}+(n+1){\bf a}+{\bf r}_1,\sigma}
- \hat c^{\dagger}_{{\bf j}+(n+1){\bf a}+{\bf r}_6,\sigma}).
\label{c2}
\end{eqnarray}
Given by its cell homogeneous nature, the $\hat B^{\dagger}$ from (\ref{c2}) 
cannot be translated or rotated in providing
new, linearly independent $\hat B^{\dagger}$ operators. 

The following linearly 
independent solutions emerge for different $A$ values from which one presents below
the $A=\sqrt{3}$ case. For this parameter region,  
$\hat B^{\dagger}_{2,\sigma}$ has the coefficients
\begin{eqnarray}
&&x_{2,1,1}=\frac{1+\sqrt{3}}{2}, x_{2,1,2}=-\frac{1+\sqrt{3}}{2g}, 
x_{2,1,3}=-\frac{1+\sqrt{3}}{g}, x_{2,1,4}=1, x_{2,1,5}=-1, 
\nonumber\\
&&\hspace*{1cm}x_{2,1,6}=-\frac{1+\sqrt{3}}{2},
x_{2,1,7}=\frac{3+\sqrt{3}}{2g}, x_{2,1,8}= \frac{3+\sqrt{3}}{2g},
\nonumber\\
&&x_{2,2,1}= \sqrt{3}, x_{2,2,2}=-\frac{1+\sqrt{3}}{g}, 
x_{2,2,3}=-\frac{1+\sqrt{3}}{2g},
x_{2,2,4}=\frac{3-\sqrt{3}}{2}, x_{2,2,5}=-\frac{3-\sqrt{3}}{2},
\nonumber\\
&&\hspace*{1cm}x_{2,2,6}=-\sqrt{3}, x_{2,2,7}=\frac{3+\sqrt{3}}{2g}, x_{2,2,8}=0,
\nonumber\\
&&x_{2,3,1}=-(1-\sqrt{3}), x_{2,3,2}=-\frac{1+\sqrt{3}}{2g}, 
x_{2,3,3}=\frac{1+\sqrt{3}}{2g}, x_{2,3,4}=1-\sqrt{3}, x_{2,3,5}=-(1-\sqrt{3}),
\nonumber\\
&&\hspace*{1cm}x_{2,3,6}=1-\sqrt{3}, x_{2,3,7}=0, x_{2,3,8}=-\frac{3+\sqrt{3}}{2g},
\nonumber\\
&&x_{2,4,1}=-\frac{3-\sqrt{3}}{2}, x_{2,4,2}=\frac{1+\sqrt{3}}{2g}, 
x_{2,4,3}=\frac{1+\sqrt{3}}{g}, x_{2,4,4}=-\sqrt{3}, x_{2,4,5}=\sqrt{3},
\nonumber\\
&&\hspace*{1cm}x_{2,4,6}=\frac{3-\sqrt{3}}{2}, x_{2,4,7}=-\frac{3+\sqrt{3}}{2g},
x_{2,4,8}=-\frac{3+\sqrt{3}}{2g},
\nonumber\\
&&x_{2,5,1}=-1, x_{2,5,2}=\frac{1+\sqrt{3}}{g}, x_{2,5,3}=\frac{1+\sqrt{3}}{2g},
x_{2,5,4}=-\frac{1+\sqrt{3}}{2}, x_{2,5,5}=\frac{1+\sqrt{3}}{2},
\nonumber\\
&&\hspace*{1cm}x_{2,5,6}=1, x_{2,5,7}=-\frac{3+\sqrt{3}}{2g}, x_{2,5,8}=0,
\nonumber\\
&&x_{2,6,1}=0, x_{2,6,2}=\frac{1+\sqrt{3}}{2g}, x_{2,6,3}=-\frac{1+\sqrt{3}}{2g},
x_{2,6,4}=0, x_{2,6,5}=0,
\nonumber\\
&&\hspace*{1cm}x_{2,6,6}=0, x_{2,6,7}=0, x_{2,6,8}=\frac{3+\sqrt{3}}{2g},
\label{c3}
\end{eqnarray}
where $g=6t'_1/(t_1+2t'_2)$ is arbitrary. This solution can be translated 
five times by
${\bf a}$ obtaining $\hat B^{\dagger}_{3,\sigma},\hat B^{\dagger}_{4,\sigma},...,
\hat B^{\dagger}_{7,\sigma}$, and the six operators $\hat B^{\dagger}_{2,\sigma},...,
\hat B^{\dagger}_{7,\sigma}$ can be rotated by $\pi$ obtaining 
$\hat B^{\dagger}_{8,\sigma},
...,\hat B^{\dagger}_{13,\sigma}$. Consequently, at this moment, the number of 
linearly
independent $\hat B^{\dagger}$ operators with fixed spin at our disposal is $13$.

The following linearly independent solution holding the index $\ell=14$ has the  
prefactors
\begin{eqnarray}
&&x_{14,1,1}=\sqrt{3}, x_{14,1,2}=-\frac{\sqrt{3}}{g}, 
x_{14,1,3}=-\frac{\sqrt{3}}{g}, x_{14,1,4}=2-\sqrt{3}, x_{14,1,5}=-(2-\sqrt{3}), 
\nonumber\\
&&\hspace*{1cm}x_{14,1,6}=-\sqrt{3},
x_{14,1,7}=\frac{2}{g}, x_{14,1,8}= \frac{1}{g},
\nonumber\\
&&x_{14,2,1}= -2(1-\sqrt{3}), x_{14,2,2}=-\frac{\sqrt{3}}{g}, x_{14,2,3}=0,
x_{14,2,4}=3-2\sqrt{3}, x_{14,2,5}=-(3-2\sqrt{3}),
\nonumber\\
&&\hspace*{1cm}x_{14,2,6}=2(1-\sqrt{3}), x_{14,2,7}=\frac{1}{g}, 
x_{14,2,8}=-\frac{1}{g},
\nonumber\\
&&x_{14,3,1}=-(3-2\sqrt{3}), x_{14,3,2}=0, x_{14,3,3}=\frac{\sqrt{3}}{g}, 
x_{14,3,4}=2(1-\sqrt{3}), x_{14,3,5}=-2(1-\sqrt{3}),
\nonumber\\
&&\hspace*{1cm}x_{14,3,6}=3-2\sqrt{3}, x_{14,3,7}=-\frac{1}{g}, 
x_{14,3,8}=-\frac{2}{g},
\nonumber\\
&&x_{14,4,1}=-(2-\sqrt{3}), x_{14,4,2}=\frac{\sqrt{3}}{g}, 
x_{14,4,3}=\frac{\sqrt{3}}{g}, x_{14,4,4}=-\sqrt{3}, x_{14,4,5}=\sqrt{3},
\nonumber\\
&&\hspace*{1cm}x_{14,4,6}=2-\sqrt{3}, x_{14,4,7}=-\frac{2}{g},
x_{14,4,8}=-\frac{1}{g},
\nonumber\\
&&x_{14,5,1}=0, x_{14,5,2}=\frac{\sqrt{3}}{g}, x_{14,5,3}=0,
x_{14,5,4}=-1, x_{14,5,5}=1,
\nonumber\\
&&\hspace*{1cm}x_{14,5,6}=0, x_{14,5,7}=-\frac{1}{g}, x_{14,5,8}=\frac{1}{g},
\nonumber\\
&&x_{14,6,1}=1, x_{14,6,2}=0, x_{14,6,3}=-\frac{\sqrt{3}}{g},
x_{14,6,4}=0, x_{14,6,5}=0,
\nonumber\\
&&\hspace*{1cm}x_{14,6,6}=-1, x_{14,6,7}=\frac{1}{g}, 
x_{14,6,8}=\frac{2}{g}.
\label{c4}
\end{eqnarray}

Similarly, another linearly independent $\hat B^{\dagger}_{\ell,\sigma}$
term $\ell=15$ is the following one
\begin{eqnarray}
&&x_{15,1,1}=\frac{3-\sqrt{3}}{2}, x_{15,1,2}=-\frac{1-\sqrt{3}}{2g}, 
x_{15,1,3}=-\frac{1}{g}, x_{15,1,4}=-(2-\sqrt{3}), x_{15,1,5}=2-\sqrt{3}, 
\nonumber\\
&&\hspace*{1cm}x_{15,1,6}=-\frac{3-\sqrt{3}}{2},
x_{15,1,7}=-\frac{1-\sqrt{3}}{2g}, x_{15,1,8}= \frac{1+\sqrt{3}}{2g},
\nonumber\\
&&x_{15,2,1}= 3-\sqrt{3}, x_{15,2,2}=-\frac{1}{g}, 
x_{15,2,3}=-\frac{1+\sqrt{3}}{2g}, x_{15,2,4}=-\frac{5-3\sqrt{3}}{2}, 
x_{15,2,5}=\frac{5-3\sqrt{3}}{2},
\nonumber\\
&&\hspace*{1cm}x_{15,2,6}=-(3-\sqrt{3}), x_{15,2,7}=\frac{1+\sqrt{3}}{2g}, 
x_{15,2,8}=\frac{1}{g},
\nonumber\\
&&x_{15,3,1}=3-\sqrt{3}, x_{15,3,2}=-\frac{1+\sqrt{3}}{2g}, 
x_{15,3,3}=\frac{1-\sqrt{3}}{2g}, x_{15,3,4}=-(2-\sqrt{3}), 
x_{15,3,5}=2-\sqrt{3},
\nonumber\\
&&\hspace*{1cm}x_{15,3,6}=-(3-\sqrt{3}), x_{15,3,7}=\frac{1}{g}, 
x_{15,3,8}=\frac{1-\sqrt{3}}{2g},
\nonumber\\
&&x_{15,4,1}=\frac{3-\sqrt{3}}{2}, x_{15,4,2}=\frac{1-\sqrt{3}}{2g}, 
x_{15,4,3}=\frac{1}{g}, x_{15,4,4}=-1, x_{15,4,5}=1,
\nonumber\\
&&\hspace*{1cm}x_{15,4,6}=-\frac{3-\sqrt{3}}{2}, x_{15,4,7}=\frac{1-\sqrt{3}}{2g},
x_{15,4,8}=-\frac{1+\sqrt{3}}{2g},
\nonumber\\
&&x_{15,5,1}=0, x_{15,5,2}=\frac{1}{g}, x_{15,5,3}=\frac{1+\sqrt{3}}{2g},
x_{15,5,4}=-\frac{1+\sqrt{3}}{2}, x_{15,5,5}=\frac{1+\sqrt{3}}{2},
\nonumber\\
&&\hspace*{1cm}x_{15,5,6}=0, x_{15,5,7}=-\frac{1+\sqrt{3}}{2g}, 
x_{15,5,8}=-\frac{1}{g},
\nonumber\\
&&x_{15,6,1}=0, x_{15,6,2}=\frac{1+\sqrt{3}}{2g}, x_{15,6,3}=-\frac{1-\sqrt{3}}{2g},
x_{15,6,4}=-1, x_{15,6,5}=1,
\nonumber\\
&&\hspace*{1cm}x_{15,6,6}=0, x_{15,6,7}=-\frac{1}{g}, 
x_{15,6,8}=-\frac{1-\sqrt{3}}{2g}.
\label{c5}
\end{eqnarray}
Both $\hat B^{\dagger}_{14,\sigma}$, and $\hat B^{\dagger}_{15,\sigma}$ 
can be translated and rotated in providing new $\hat B^{\dagger}$ operators 
in the ground state wave vector, but from these not all 
are linearly independent. 

One further note that maintaining (\ref{c1}), solutions can be found also for
other $A$ values, as for example $A=0.75, 1, 2, 4, \sqrt{2}, (\sqrt{5} \pm 1)/2,$
etc. Furthermore, solutions are present even if (\ref{c1}) 
does not hold, but in this 
case cell homogeneous $\hat B^{\dagger}$ operators as presented in (\ref{c2}), are not
present.

%%%%%%%%%%%%%%%%%%%%%%%%%%%%%%%%%%%%%%%%%%%%%%%%%%%%%%%%%%%%%%%%%%%%%%%%%%%%%%%%%%%% 
%\begin{thebibliography}

%\end{thebibliography}

\end{document}